\documentclass[a4paper,fleqn,usenatbib]{mnras}
\usepackage[T1]{fontenc}
\usepackage{ae,aecompl}

\usepackage{amsmath}
\usepackage{graphicx} 
\usepackage{comment}
\usepackage{bm}

\def\aj{AJ}%
\def\apj{ApJ}%
\def\apjl{ApJ}%
\def\apjs{ApJS}%
\def\ao{Appl.~Opt.}%
\def\aap{A\&A}%
\def\mnras{MNRAS}%
\def\prd{Phys.~Rev.~D}%
\def\nat{Nature}%
\title[Combined constraints on MOND]{Combined Solar System and rotation curve constraints on MOND} 
\author[A. Hees et al.]  {
Aur\'elien Hees$^{1}$\thanks{A.Hees@ru.ac.za}, Benoit Famaey$^2$, Garry W. Angus$^3$ and Gianfranco Gentile$^{3,4}$ \vspace{10pt} \\
$^1$ Department of Mathematics, Rhodes University, 6140 Grahamstown, South Africa \\
$^2$ Observatoire astronomique de Strasbourg, Universit\'e de Strasbourg, CNRS, UMR 7550, 11 rue de l'Universit\'e, F-67000 Strasbourg, France \\
$^3$ Department of Astronomy and Astrophysics, Vrije Universiteit Brussel, Pleinlaan 2, 1050 Brussels, Belgium \\
$^4$ Sterrenkundig Observatorium, Universiteit Gent, Krijgslaan 281, 9000, Gent, Belgium
}

\pubyear{2015}

\begin{document}
\label{firstpage}
\pagerange{\pageref{firstpage}--\pageref{lastpage}}
\maketitle

\begin{abstract}
The Modified Newtonian Dynamics (MOND) paradigm generically predicts that the external gravitational field in which a system is embedded can produce effects on its internal dynamics. In this communication, we first show that this External Field Effect can significantly improve some galactic rotation curves fits by decreasing the predicted	 velocities of the external part of the rotation curves. In modified gravity versions of MOND, this External Field Effect also appears in the Solar System and leads to a very good way to constrain the transition function of the theory. A combined analysis of the galactic rotation curves and Solar System constraints (provided by the Cassini spacecraft) rules out several classes of popular MOND transition functions, but leaves others viable. Moreover, we show that LISA Pathfinder will not be able to improve the current constraints on these still viable transition functions.
\end{abstract}


\section{Introduction}

With only six free parameters, the standard $\Lambda$CDM cosmological model fits no less than 2500 multipoles in the Cosmic Microwave Background (CMB) angular power spectrum~\citep{planck-collaboration:2014yf}, the Hubble diagram of Type Ia supernovae, the large-scale structure matter power spectrum, and even the detailed scale of baryonic acoustic oscillations. It thus provides the current basis for simulations of structure formation, and is extremely successful down to the scale of galaxy clusters and groups. Nevertheless, it still faces numerous challenges on galaxy scales. Among these, the most important ones are the too-big-to-fail problem~\citep{boylan-kolchin:2011kq} and the satellite-plane problem~\citep[e.g.][]{pawlowski:2012qf,ibata:2014xy} for dwarf galaxies, the tightness of the baryonic Tully-Fisher relation~\citep{mcgaugh:2012nr,vogelsberger:2014sf}, or the unexpected diversity of rotation curve shapes at a given mass-scale~\citep{oman:2015rm}. The latter problem is actually a subset of a more general problem, i.e. that the shapes of rotation curves indeed do not depend on the DM halo mass, contrary to what would be expected in $\Lambda$CDM, but rather on the baryonic surface density, as has long been noted~\citep[e.g.,][]{zwaan:1995yg}. This makes the problem even worse, since the rotation curve shapes are not only diverse at a given mass-scale, but uniform at a given baryonic surface density scale, implying a completely ununderstood fine-tuning of putative feedback mechanisms. On the other hand, this behaviour of rotation curves is an {\it a priori} prediction of the formula proposed by Milgrom more than 30 years ago~\citep{milgrom:1983uq,milgrom:1983fk}, relating the total gravitational field to the Newtonian field generated by baryons alone, and which can be interpreted as a modification of Newtonian dynamics on galaxy scales below a characteristic acceleration~\citep[MOND, for a review see][]{famaey:2012fk,milgrom:2014ix}. With this simple formula, High Surface Brightness  (HSB) galaxies are predicted to have rotation curves that rise steeply before becoming essentially flat, or even falling somewhat to the not-yet-reached asymptotic circular velocity, while Low Surface Brightness (LSB) galaxies are predicted to have rotation curves that rise slowly to the asymptotic velocity. This is precisely what is observed, and was predicted by Milgrom long before LSB galaxies were even known to exist. The formula also predicts the tightness of the baryonic Tully-Fisher relation.

Since the original formulation of the MOND paradigm, a lot of relativistic theories of gravitation reproducing the MOND regime in very weak fields  have been developed. Usually, these GR extensions imply the presence of additional scalar or vector fields in addition to the standard metric to mediate the gravitational interaction. These relativistic MOND theories include the original Bekensetein tensor-vector-scalar (TeVeS) theory~\citep{bekenstein:2004fk,sanders:1997kx,sanders:2005fk}, Einstein-Aether theories~\citep{jacobson:2001qf,zlosnik:2006dq,zlosnik:2007bh}, bimetric theories~\citep{milgrom:2009kx}, or non-local theories~\citep{deffayet:2014kx}. Reviews of the relativistic extensions of the MOND paradigm can be found in \citet{bruneton:2007vn} and in \citet{famaey:2012fk}. More recently, new interpretations of MOND in terms of a non-standard DM fluid have been developed~\citep{blanchet:2007uq,blanchet:2008fv,blanchet:2009kx,bernard:2015sf,blanchet:2015rm,khoury:2015qf,berezhiani:2015bh}, in which case Milgrom's formula is akin to an {\it effective} modification of gravity on galaxy scales. These latter theories have the advantage of naturally reproducing the CMB power spectrum, and to basically differ from $\Lambda$CDM only on galaxy scales and below.

In the non-relativistic regime on galaxy scales and below, almost all\footnote{For instance, in the case of non-standard DM theories reproducing MOND, this can nevertheless depend on the presence or not of the DM fluid in the systems under consideration.} these theories boil down to two types of modified Poisson equations, which we explicitly discuss in Sec.~2. One feature of MOND is that it generically (at least for all modified gravity theories) predicts a violation of the strong equivalence principle. This implies that the internal gravitational dynamics of a system depends on the external gravitational field in which the system is embedded~\citep{milgrom:1983fk}. This External Field Effect (EFE) occurs even for a constant external gravitational field\footnote{Of course, if the external field is not constant, it will produce additional standard tidal effects.}, and it can have observational effects, in particular for computing the escape speed from galaxies~\citep{famaey:2007dq,wu:2008cr}, in the rotation curve of the outskirts of galaxies, and even in the Solar System~\citep{milgrom:2009vn,blanchet:2011ys}. The latter can put stringent constraints on the transition behaviour between the high-acceleration Newtonian regime and the low-acceleration MOND regime, which we investigate in details in the present contribution. Another question is whether deviations from General Relativity could be detected close to the saddle point of the gravitational potential in the Solar System~\citep[e.g.][]{bekenstein:2006sf}, thereby putting additional constraints on MOND. Here we check in particular whether measurements from the LISA pathfinder mission could add new constraints to existent ones from other Solar System tests.

In Sec.~\ref{sec:basics}, we review the basics of MOND, in Sec.~\ref{sec:rotation} we produce rotation curve fits to a sample of galaxies with various transition functions, including for the first time the EFE in the fits, in Sec.~\ref{sec:solsys} we combine the best-fit values of the rotation curve MOND fits with existing Solar System constraints to exclude a large range of transition functions, and check whether improved constraints could be obtained with LISA pathfinder. We conclude in Sec.~\ref{sec:conclusion}.

\section{MOND basics}\label{sec:basics}
The original idea of the MOND paradigm is to modify the standard Newtonian gravitation law $\bm a=\bm g_N$ (where $\bm a$ is the acceleration of a body and $\bm g_N$ is the Newtonian gravitational field) by the relation $\bm a=\bm g$ with $\bm g$ determined by the relation
\begin{equation}\label{eq:mua}
 \mu\left(\frac{g}{a_0}\right)\bm g=\bm g_N \, .
\end{equation}
or
\begin{equation}\label{eq:nua}
 \nu\left(\frac{g_N}{a_0}\right)\bm g_N=\bm g \, .
\end{equation}
In these expressions, $\mu$ or $\nu$ is the MOND interpolating function or transition function. The MOND regime appears in weak gravitational fields ($g<<a_0$) where the transition function needs to satisfy $\mu(x)\rightarrow x$ or $\nu(y) \rightarrow y^{-1/2}$ in order to explain the galactic rotation curves~\citep{milgrom:1983fk,milgrom:1983uq}. On the other hand, in order to recover the very well constrained Newtonian regime in the Solar System, the MOND transition function has to satisfy $\mu(x)\rightarrow 1$ or $\nu(y)\rightarrow 1$ for $g>>a_0$.

An equation such as Eq.~(\ref{eq:mua}) or Eq.~(\ref{eq:nua}) cannot be valid outside of spherical symmetry for any type of orbit~\citep{felten:1984uq}. A first approach for a more fundamental underlying theory is known as Modified Inertia. implying that the particle equations of motion are modified while the gravi- tational potential is still given by the standard Newtonian potential~\citep{milgrom:1994zr,milgrom:2011fk}. These theories are typically nonlocal  and Eq.~(\ref{eq:mua}) or Eq.~(\ref{eq:nua}) is then valid only for circular orbits.

All relativistic theories of MOND are rather Modified Gravity theories (or effective modified gravity in the case of non-standard DM), and in the non-relativistic regime they basically reduce to two types of modified Poisson equation:
\begin{itemize}	
	\item The first one takes the non-linear form~\citep{bekenstein:1984kx}
	\begin{equation}\label{eq:MONDBek}
		\bm\nabla.\left[\mu\left(\frac{\left|\bm\nabla \Phi\right|}{a_0}\right)\bm\nabla \Phi  \right]= 4\pi G \rho=\bm \nabla^2 \Phi_N \, ,
	\end{equation}
	with $G$ the Newtonian constant, $\rho$ the matter density, $\Phi_N$ the Newtonian gravitational potential solution of the standard Poisson equation. The gravitational potential $\Phi$ is the MONDian gravitational potential that enters the particle's equations of motion $\bm a=-\bm\nabla\Phi$. This is typically the weak-field limit of MOND-inspired Einstein-Aether theories~\citep{zlosnik:2007bh}.
	
	\item The second is called quasi-linear MOND (or QUMOND)~\citep{milgrom:2010uq}. In QUMOND, the gravitational field is the solution of the equation
	\begin{equation}\label{eq:QUMOND}
		\bm \nabla^2\Phi =\bm \nabla . \left[\nu\left(\frac{\left|\bm\nabla \Phi_N\right|}{a_0}\right)\bm \nabla\Phi_N\right]\, .
	\end{equation}
	This approach requires solving two linear Poisson equations to find the gravitational potential $\Phi$ (for the previous approach, we had to solve a non-linear Poisson equation). This can be the weak-field limit of bimetric MOND theories~\citep{milgrom:2009kx}.
\end{itemize}
It is known that these two equations are fully equivalent in  spherically symmetric situations~\citep{milgrom:2010uq,zhao:2010uq}. In that case, the transition functions $\mu$ and $\nu$ are related by $\nu(y)=1/\mu(x)$ with $x$ and $y$ related through $x\mu(x)=y$~\citep{milgrom:2010uq}.

\begin{figure}
\includegraphics[width=.9\columnwidth]{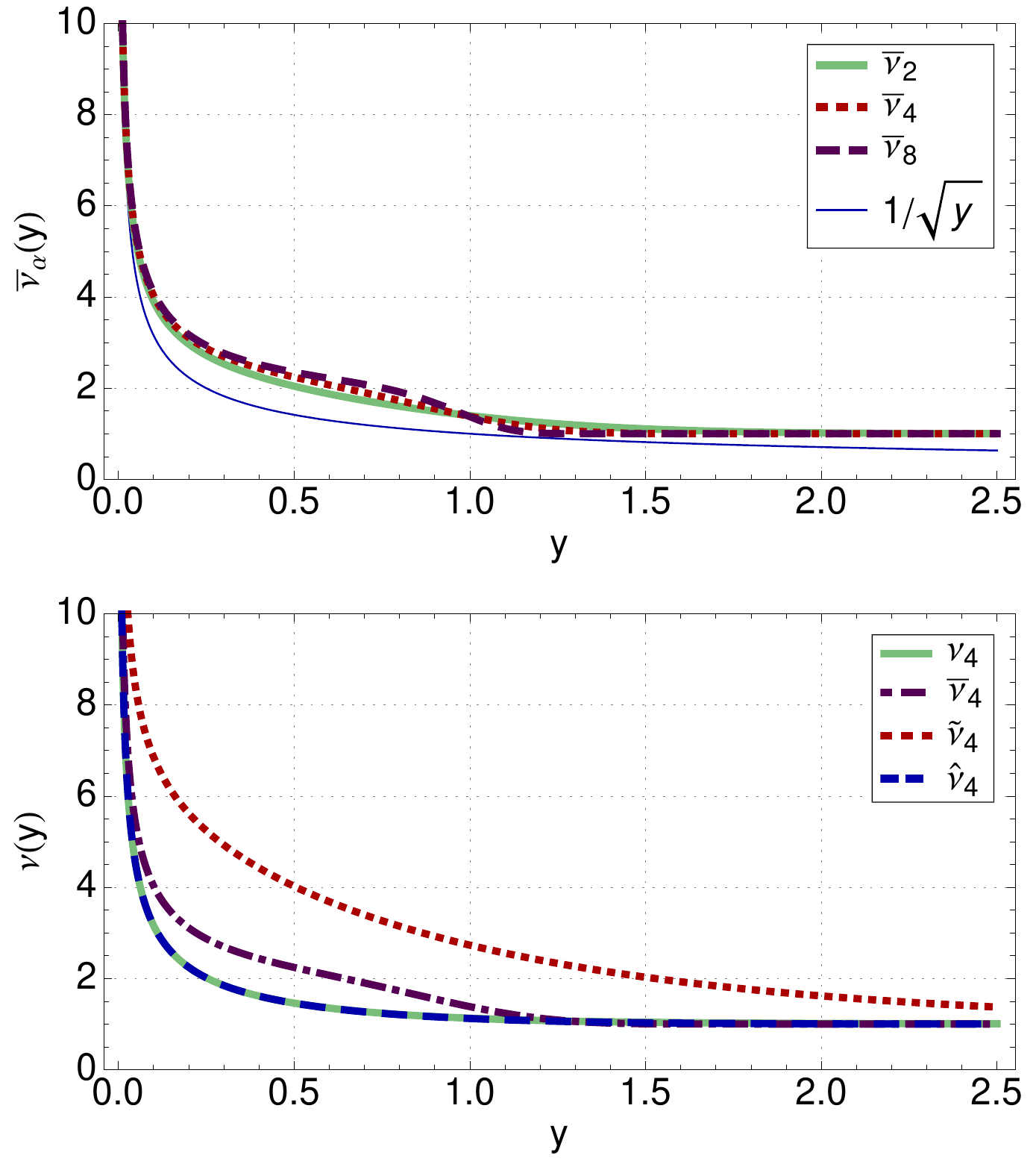}
\caption{Representation of different MOND transition functions $\nu$ (see Eqs.~(\ref{eq:nu_family}) for their expression).}
\label{fig:nu}
\end{figure}

Different types of MOND transition function have been used in the literature, the most common families of functions being~\citep{famaey:2012fk}
\begin{subequations}\label{eq:nu_family}
	\begin{eqnarray}
		\nu_\alpha(y)&=&\left[\frac{1+\left(1+4y^{-\alpha}\right)^{1/2}}{2}\right]^{1/\alpha}\, ,  \label{eq:nun}\\
		\tilde \nu_\alpha(y)&=&\left(1-e^{-y}\right)^{-1/2}+\alpha \, e^{-y}\, ,\\
		\bar \nu_\alpha(y)&=&\left(1-e^{-y^\alpha}\right)^{-1/2\alpha}+\left(1-1/2\alpha\right)e^{-y^\alpha} \, ,\\
		\hat \nu_\alpha(y)&=&\left(1-e^{-y^{\alpha/2}}\right)^{-1/\alpha}\, .
	\end{eqnarray}
\end{subequations}
For instance, $\nu_1$ is the so-called ``simple" interpolating function~\citep{famaey:2005rq,zhao:2006la}, $\nu_2$ is the ``standard" one, and $\bar\nu_{0.5}$ has been extensively used in \citet{famaey:2012fk}. Fig.~\ref{fig:nu} represents all these different transition functions. The family of functions $\bar\nu_\alpha$ is presented for different values of $\alpha$ as we will see in Sec.~\ref{sec:solsys} that this family is the most promising one to fit rotation curves and to satisfy Solar System constraints simultaneously.

The EFE mentioned in the previous section is due to the fact that the MOND equations (\ref{eq:MONDBek}) and (\ref{eq:QUMOND}) are non-linear and involve the total gravitational acceleration with respect to a pre-defined frame (e.g. the CMB frame). Decomposing the total gravitational field $\bm\nabla\Phi$ into an internal part $\bm g$ and an external field $\bm g_e$ and using a similar decomposition for the Newtonian gravitational acceleration ($\bm \nabla \bm \Phi_N=\bm g_N+\bm g_{Ne}$) allows us to solve the equations by taking into account the external field. This must typically be done with a numerical Poisson solver~\citep{wu:2008cr,angus:2012cs,lughausen:2015jt}. Nevertheless, fits to rotation curves in MOND usually neglect the small corrections due to the non-spherical symmetry of the problem, in order to allow for a direct fit of the rotation curve. In the same spirit, and in order to get a first glimpse of the influence of the EFE on rotation curves, we generalize the one-dimensional solution, by using the following formula to fit rotation curves, namely Eq.(60) from \citet{famaey:2012fk}:
\begin{equation}\label{eq:EFE}
      \bm g=\nu\left(\frac{|\bm g_N+\bm g_{Ne}|}{a_0}\right)\left(\bm g_N + \bm g_{Ne}\right)-\nu\left(\frac{g_{Ne}}{a_0}\right)\bm g_{Ne}\, .
\end{equation}
The 1D version of this formula has been shown to be a good approximation of the true 3D solution from a numerical Poisson solver for a random orientation of the external field, at least for computing the Galactic escape speed~\citep{famaey:2007dq,wu:2008cr}. Further work should investigate the range of variation of the actual rotation curve compared to the one obtained in this way, for full numerical solutions of the modified Poisson equation and various orientations of the EFE. As mentioned in~\citet{famaey:2012fk}, the EFE is negligible if $g_e<<g$ but can play a significant role when the gravitational field $g\sim g_e <a_0$. This condition is always reached at some point in the external part of the galaxies. In this case, the relation (\ref{eq:EFE}) shows that the EFE will induce a decrease in the internal gravitational field. In other words, the EFE can lead to a decrease of the external part of the rotation curves. We will study this effect more carefully in Sec.~\ref{sec:rotation}.

On the other hand, the EFE  also  subtly affects  the internal dynamics of the Solar System. It has been shown that within the MOND paradigm, the external field of our galaxy produces a quadrupolar modification of the Newtonian potential \citep{milgrom:2009vn,blanchet:2011ys} which is present even in the case of a rapidly vanishing transition function. As mentioned in \citep{milgrom:2009vn,blanchet:2011ys,blanchet:2011zr,hees:2012fk}, planetary ephemerides analysis (in particular from Saturn) is sensitive to this effect. An estimation of this quadrupolar modification of the Newtonian potential has been performed using Cassini radioscience data~\citep{hees:2014jk}. We will use  this estimation here to constrain the transition functions in Sec.~\ref{sec:solsys}.

\section{Rotation curve fits}\label{sec:rotation}
In this section, we produce traditional MOND fits to rotation curves~\citep{begeman:1991fk,sanders:1998uq,de-blok:1998zr,sanders:2007ly,gentile:2011uq} using different transition functions. In particular, we determine how the best-fit value of $a_0$ changes with the adopted transition. Furthermore, the influence of the EFE on galactic rotation curves will be assessed for the first time. 

We use rotation curve data from 27 dwarf and low surface brightness galaxies, for which the MOND effect is important, and that have low-enough accelerations in the outer parts for the EFE to perhaps play a role. The dataset used is thoroughly described in \citet[][hereafter SSM10]{swaters:2010eu}. In the following, we will study the influence of the chosen MOND transition function $\nu$ and of the corresponding MOND acceleration scale $a_0$. Moreover, we will also allow some freedom on local galactic parameters: the individual $R$-band stellar mass to light (M/L) ratio $\Upsilon_g$, a rescaling of the distance to the different galaxies ($d_g$) and a hypothetical external Newtonian gravitational field $g_{Neg}$ (the indices $g$ refer to a particular galaxy and indicate that the parameters are local parameters).

As stated above, the gravitational field is given by the 1D version of Eq.~(\ref{eq:EFE}). The predicted rotation velocity is given by
\begin{equation}
	V_M(R_id_g;a_0,\Upsilon_g,d_g,g_{Neg})=\sqrt{R_i d_g g(R_i;a_0,\Upsilon_g,g_{Neg} )} \, ,
\end{equation}
where $V_M$ is the predicted MONDian velocity at radius $R_i$, $\Upsilon_g$ is the stellar M/L ratio, $d_g$ is a distance scale factor $d_g=D_{g,\textrm{MOND}}/D_{g,0}$ where $D_{g,0}$ are the distances given in Tab.~1 of SSM10 and $g_{Neg}$ is the Newtonian external field. The norm of the gravitational field $g$ is determined by Eq.~(\ref{eq:EFE}) where the Newtonian gravitational field is given by
\begin{equation}
	g_N(R_i,\Upsilon_g)=\frac{V^2_{{\rm gas}i}}{R_i}+\Upsilon_g\frac{V_{\star i}^2}{R_i} \, ,
\end{equation}
where $V_{{\rm gas}i}$ and $V_{\star i}$ are the contribution of the gas and of the stellar disk (at radius $R_i$) to the rotation curves calculated in the Newtonian regime. In what precedes, we have used the fact that the Newtonian observed velocities due to the gas and to the stellar disk are rescaled as $\propto \sqrt{d}$ with a distance rescaling. Similarly, the measured radial distances $R_i$ are rescaled proportionally to $d$. The procedure then consists of the two following steps:
\begin{enumerate}
	\item Step 1: Using a subset of 19 galaxies from SSM10, we perform a least-squares fit of the global MOND acceleration scale $a_0$ and of the local $\Upsilon_g$ and $d_g$ parameters neglecting the external field $g_{Neg}=0$. The galaxies not considered in this part of the analysis are the galaxies that seem to experience a potentially non-negligible EFE, i.e. where the MOND fit is slightly too large for the external part of the rotation curves (this first MOND fit was actually made in a previous step, Step 0, where all galaxies are taken into account). The goal of this first step is to find a robust estimation of the MOND acceleration scale $a_0$ that is not influenced by the EFE.
	
	\item Step 2: Using the optimal value of $a_0$ obtained from the first step, we perform a local fit of the parameters $\Upsilon_g$, $d_g$ and $g_{Neg}$ for each of the 27 galaxies from the dataset of SSM10. This fit is done using a standard Bayesian inversion with a Metropolis-Hasting Monte Carlo Markov Chain (MCMC) algorithm~\citep{gregory:2010qv}. The marginalized posterior distribution of the parameter $g_{Neg}$ allows us to identify  the galaxies with a significantly non-vanishing external field.
	
\end{enumerate}

During the analysis, we always impose a constraint that the stellar M/L ratios must have values included between 0.3 and 5 (in units of $(M/L)_{\sun}$). Similarly, we require the scaling of the distance $d$ to be  between 0.7 and 1.3 which corresponds to the standard uncertainties on the distances~\citep{swaters:2002qf}. Furthermore, we also include a Gaussian prior (characterized by a mean of 1 and a standard deviation of 0.1) on the parameters $d_g$. In this analysis, we consider a large range of MOND transition functions from all the families $\nu_\alpha$, $\bar \nu_\alpha$, $\hat \nu_\alpha$ and $\tilde \nu_\alpha$. 

\subsection{Global fit of the MOND acceleration using a subset of the dataset}
\begin{figure*}
\includegraphics[width=1.8\columnwidth]{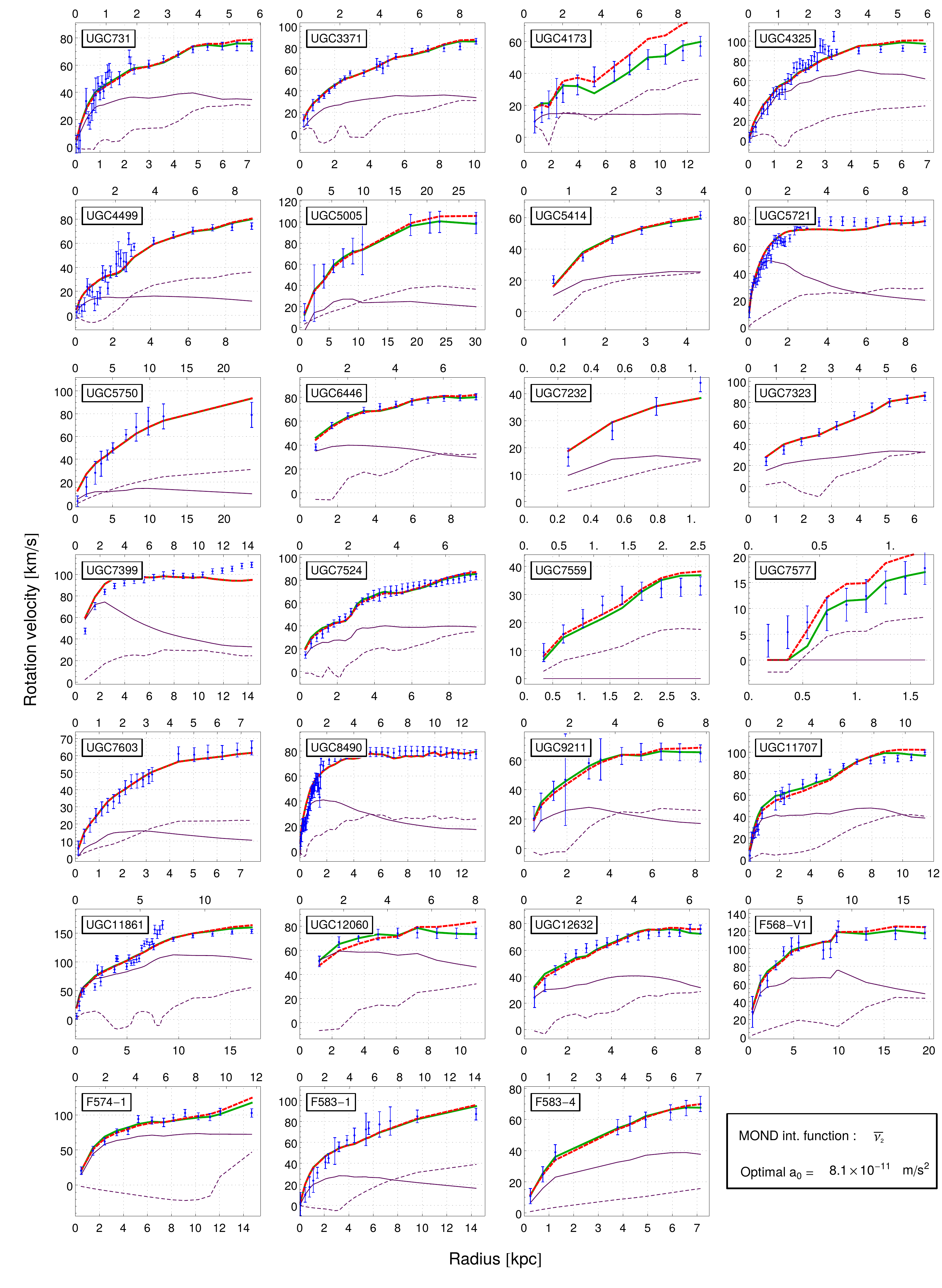}
\caption{Results of the fit using the MOND transition function $\bar\nu_2$, which is compatible with Solar System constraints (see Sec.~4), for the optimal value $a_0=8.1\times 10^{-11}$ m/s$^2$. The dashed (red) thick lines represent the optimal fit without any EFE ; the thick (green) solid line represents the optimal fit with EFE ; the thin solid line represents the Newtonian contributions of the stars and the thin dashed line represents the gas contribution. Since the optimal fits with and without EFE do not necessarily produce the same distance scale factor, the radial scales may not be the same. On the top of the plots we mention the radial scale obtained without EFE (corresponding to the dashed red thick lines), at the bottom of the plots we mention the radial scale obtained with EFE (corresponding to the thick green solid lines).}
\label{fig:resnu2}
\end{figure*}

We perform a global fit of the global MOND acceleration scale $a_0$ and the local parameters $\Upsilon_g$ and $d_g$ using all the 27 galaxies (Step 0). The EFE is neglected at this stage. For the function $\bar \nu_2$, which we take as a representative example throughout this analysis, this first global fit leads to an optimal value of $a_0=7.5\times 10^{-11}$ m/s$^2$. The purpose of this first fit is only to identify which galaxies seem to experience a non negligible EFE. We identify these galaxies as being the ones where the MOND fit  statistically produces a too high velocity on the last points of the rotation curves. The so identified galaxies are: UGC 4173, UGC 4325, UGC 7559, UGC 7577, UGC 11707, UGC 11861, UGC 12060, F574-1. A new global fit using the 19 other galaxies leads to a new optimal value $a_0=8.1\times 10^{-11}$ m/s$^2$. This new value is more robust and less influenced by the EFE. The local optimal parameters obtained for each galaxy for this optimal MOND acceleration scale are given in Tab.~\ref{tab:resbarnu2} (let us note here again that the EFE is neglected in this first part) and the obtained rotation curves for $\bar \nu_2$ are shown in Fig.~\ref{fig:resnu2}. The same procedure is repeated for a large class of transition functions and the resulting best-fit $a_0$ for each function is presented in Tab.~\ref{tab:efe_SS}.

\begin{table*}
\caption{Best-fit parameters obtained for the MOND transition function $\bar \nu_2$ and for $a_0=8.1\times 10^{-11}$ m/s$^2$. Cols. 2 and 3 are the optimal values obtained in the case where the EFE is neglected. Cols. 4, 5 and 6 are optimal values and 68 \% Bayesian confidence intervals for the parameters when the EFE is taken into account. The values of the external gravitational field are mentioned only when different from 0.}
\label{tab:resbarnu2} 
\centering
\begin{tabular}{l|rr|rrr}
\hline
Name &  \multicolumn{2}{|c}{No EFE} & \multicolumn{3}{|c}{With EFE}\\
& $\Upsilon_g$ & $d_g$ &  $\Upsilon_g$ & $d_g$ & $\log g_{ep}$ \\
&$(M/L)_{\sun}$&&$(M/L)_{\sun}$& & [m/s$^2$]\\\hline
UGC 731  &  5.0  &   0.82  &  $5.0_{-0.5}^{+0.0}$  &  \
$1.03_{-0.19}^{+0.04}$  &  $-11.80_{-1.12}^{+0.44}$  \\ 
UGC 3371  &  3.2  &   0.86  &  $3.3_{-0.6}^{+0.4}$  &  \
$0.99_{-0.19}^{+0.00}$  &  --   \\ 
UGC 4173  &  0.3  &   0.70  &  $0.8_{-0.4}^{+0.5}$  &  \
$0.99_{-0.12}^{+0.10}$  &  $-11.20_{-0.22}^{+0.37}$  \\ 
UGC 4325  &  3.1  &   0.94  &  $3.7_{-0.6}^{+0.5}$  &  \
$1.10_{-0.10}^{+0.07}$  &  $-11.30_{-0.39}^{+0.26}$  \\ 
UGC 4499  &  0.3  &   0.97  &  $0.3_{-0.0}^{+0.0}$  & \
$1.03_{-0.11}^{+0.00}$  &  --  \\ 
UGC 5005  &  0.6  &   0.93  &  $0.9_{-0.6}^{+0.1}$  &  \
$1.01_{-0.16}^{+0.02}$  &  $-12.70_{-5.37}^{+0.76}$  \\ 
UGC 5414  &  0.6  &   0.84  &  $1.0_{-0.5}^{+0.1}$  &  \
$0.93_{-0.18}^{+0.01}$  &  $-11.90_{-4.92}^{+1.60}$  \\ 
UGC 5721  &  2.4  &   1.23  &  $2.4_{-0.2}^{+0.3}$  &  \
$1.23_{-0.04}^{+0.07}$  & --  \\ 
UGC 5750  &  0.3  &   1.02  &  $0.3_{-0.0}^{+0.2}$  &  \
$1.02_{-0.13}^{+0.04}$  &  --  \\ 
UGC 6446  &  1.8  &   0.72  &  $1.6_{-0.1}^{+0.4}$  &  \
$0.91_{-0.21}^{+0.00}$  &  $-12.20_{-1.53}^{+0.62}$  \\ 
UGC 7232  &  0.8  &   1.03  &  $0.8_{-0.3}^{+0.4}$  &  \
$1.04_{-0.12}^{+0.09}$  &  --  \\ 
UGC 7323  &  0.6  &   1.01  &  $0.6_{-0.1}^{+0.2}$  &  \
$1.01_{-0.09}^{+0.08}$  &  --  \\ 
UGC 7399  &  5.0  &   1.30  &  $5.0_{-0.1}^{+0.0}$  &  \
$1.30_{-0.01}^{+0.00}$  &  --  \\ 
UGC 7524  &  1.8  &   0.70  &  $1.9_{-0.3}^{+0.3}$  &  \
$0.91_{-0.20}^{+0.03}$  &  $-11.70_{-0.72}^{+0.47}$  \\ 
UGC 7559  &  0.0  &   0.79  &  $0.0_{-0.4}^{+0.6}$  &  \
$0.96_{-0.22}^{+0.00}$  &  $-12.50_{-4.27}^{+0.61}$  \\ 
UGC 7577  &  0.0  &   0.76  &  $0.0_{-0.4}^{+0.6}$  &  \
$1.00_{-0.11}^{+0.11}$  &  $-12.00_{-0.29}^{+0.57}$  \\ 
UGC 7603  &  0.4  &   1.17  &  $0.4_{-0.1}^{+0.1}$  &  \
$1.17_{-0.08}^{+0.05}$  &  --  \\ 
UGC 8490  &  1.4  &   1.30  &  $1.4_{0.0}^{+0.7}$  &  \
$1.30_{-0.14}^{+0.00}$  & --  \\ 
UGC 9211  &  2.3  &   0.94  &  $3.0_{-1.4}^{+0.6}$  &  \
$1.00_{-0.14}^{+0.03}$  &  $-12.70_{-5.78}^{+0.00}$  \\ 
UGC 11707  &  2.6  &   0.70  &  $3.9_{-0.8}^{+0.3}$  &  \
$0.71_{-0.01}^{+0.10}$  &  $-12.10_{-0.13}^{+0.60}$  \\ 
UGC 11861  &  2.4  &   0.77  &  $2.5_{-0.3}^{+0.3}$  &  \
$0.97_{-0.13}^{+0.06}$  &  $-11.30_{-0.39}^{+0.26}$  \\ 
UGC 12060  &  2.8  &   0.73  &  $4.9_{-1.3}^{+0.1}$  &  \
$1.00_{-0.11}^{+0.09}$  &  $-10.80_{-0.47}^{+0.06}$  \\ 
UGC 12632  &  4.7  &   0.75  &  $5.0_{-0.8}^{+0.0}$  &  \
$0.95_{-0.17}^{+0.07}$  &  $-11.80_{-0.99}^{+0.38}$  \\ 
F568-V1  &  4.9  &   0.91  &  $5.0_{-0.8}^{+0.0}$  &  \
$1.04_{-0.13}^{+0.03}$  &  $-12.10_{-5.18}^{+0.45}$  \\ 
F574-1  &  3.7  &   0.78  &  $5.0_{-0.8}^{+0.0}$  &  \
$0.98_{-0.17}^{+0.05}$  &  $-11.10_{-0.81}^{+0.21}$  \\ 
F583-1  &  2.3  &   0.95  &  $2.3_{-0.5}^{+0.5}$  &  \
$0.97_{-0.11}^{+0.00}$  &  --  \\ 
F583-4  &  1.7  &   0.99  &  $3.3_{-2.0}^{+0.0}$  &  \
$1.00_{-0.13}^{+0.08}$  &  $-11.70_{-6.46}^{+0.76}$  \\ 
\hline
\end{tabular}
\end{table*}

\subsection{Local fits with the EFE}
In the second step, we use the fixed value of $a_0$ obtained previously (i.e. $a_0=8.1\times 10^{-11}$ m/s$^2$ for $\bar\nu_2$) and we perform local fits of $\Upsilon_g$, $d_g$ and $g_{Neg}$ for each of the 27 galaxies from the dataset. This part of the analysis is performed using a MCMC algorithm. Let us remind the reader that we use a flat prior on $M/L$ between 0.3 and 5. Moreover, a Gaussian prior is used on $d_g$ (with mean 1 and standard deviation 0.1). In addition, we force $d_g$ to have a value included between 0.7 and 1.3. This approach allows us to find realistic confidence intervals for the three parameters and to assess correlations.

\begin{figure}
\includegraphics[width=.9\columnwidth]{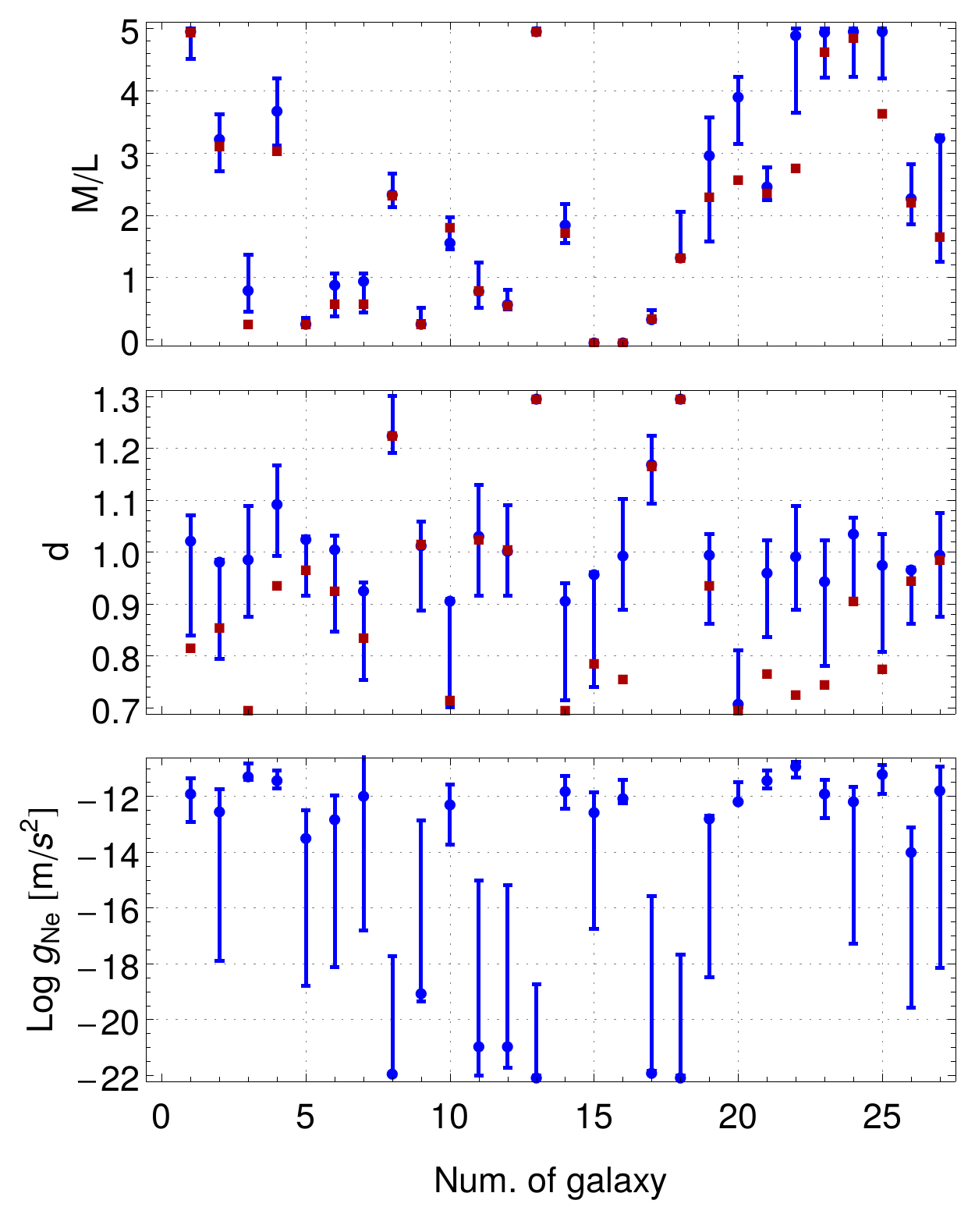}
\caption{Blue: 68 \% Bayesian confidence intervals obtained with local fits on each of the 27 galaxies with the transition function $\bar\nu_2$ and $a_0=8.1\times 10^{-11}$ m/s$^2$. The blue dots represent the best fit. The red squares represent the optimal values obtained without any external field effect.}
\label{fig:conf}
\end{figure}

The  marginalized 68 \% Bayesian confidence intervals on the parameters are presented in Tab.~\ref{tab:resbarnu2} and represented in Fig.~\ref{fig:conf}. For the EFE parameters, we only present the estimations that produce a non-vanishing $g_{Neg}$. The obtained rotation curves are also presented in Fig.~\ref{fig:resnu2}. In addition, this analysis shows a correlation between the $d_g$ and the $\Upsilon_g$ parameters. A higher estimation of the M/L ratio will lead to a lower estimation of the distance ratio. Moreover, as can be seen from Fig.~\ref{fig:conf}, taking into account the EFE produces estimations of $d$ that are slightly higher, while the estimations of the stellar M/L ratios do not change significantly.

The EFE improves spectacularly a few of the rotation curves fits. In particular, rotation curves for galaxies UGC4173 and UGC7577 are now very well fitted whereas the quality of the fit was quite poor when neglecting the EFE. The case of the galaxy UGC12060 is also very interesting since the fit is also improved  at low radii because of an increase of the inner part of the rotation curves, which is mainly due to an increase of the optimal M/L. A similar situation is encountered for UGC 11707. The fits of UGC 731, UGC5005, UGC 7559, UGC 9211, F568-V1, F574-1 are slightly improved by the addition of the EFE.

In Appendix~\ref{app:fit}, in addition to $\bar\nu_2$, we present fits using two other transition functions: $\nu_8$ and $\hat\nu_6$. We will show in the next section that these functions ($\bar\nu_2$, $\nu_8$ and $\hat\nu_6$), while produce good fits to rotation curves, are not rejected by Solar System constraints.

\section{Solar System constraints}\label{sec:solsys}
\subsection{The Solar System EFE and its constraint using Cassini data}
As shown in the previous section, the MOND EFE can have a non negligible effect on the outer parts of some galaxy rotation curves. This effect turns out to be crucial within the Solar System. Indeed, within the MOND paradigm, the external gravitational field of our galaxy produces interesting modifications in the internal dynamics of the Solar System~\citep{milgrom:2009vn,blanchet:2011ys}. The main effect consists in a quadrupole correction to the Newtonian potential. This correction can phenomenologically be parametrized by $Q_2$ and the gravitational potential can be written as
\begin{equation}\label{eq:EFE_SS}
 \Phi=-\frac{GM}{r}-\frac{Q_2}{2}x^i x^j \left(e_i e_j-\frac{1}{3}\delta_{ij}\right)\, ,
\end{equation}
where ${\bm e}$ is a unitary vector pointing towards the Galactic center and $-GM/r$ is the standard Newtonian potential due to the Sun. This correction produces an anomalous force which, along the Galactic external field direction, rises linearly with distance from the Sun, whilst it decreases linearly along the two other cartesian axes (for a positive value of $Q_2$). From a theoretical point of view, the value of $Q_2$ depends on the MOND transition function, on the value of the external gravitational field $g_e$ and on the value of the MOND acceleration scale $a_0$.  Instead of working with $Q_2$, one can introduce a dimensionless parameter $q$ defined by~\citep{milgrom:2009vn}
\begin{equation}\label{eq:q_q2}
 q=-\frac{2Q_2(GM)^{1/2}}{3 a_0^{3/2}}\, .
\end{equation}
This dimensionless parameter depends only on the MOND transition function and on the ratio
\begin{equation}
 \eta=\frac{g_e}{a_0}\, 
\end{equation}
between the external field and the MOND acceleration scale.

In the context of the QUMOND formulation (\ref{eq:QUMOND}), \citet{milgrom:2009vn} has derived an exact expression for the $q$ parameter given by
\begin{equation}\label{eq:q}
 q(\eta)=\frac{3}{2}\int_0^\infty dv\int_{-1}^1 d\xi \left(\nu-1\right) \left[\eta_N(3\xi-5\xi^3)+v^2(1-3\xi^2)\right] \, ,
\end{equation}
with $\nu=\nu\left[\sqrt{\eta_N^2+v^4+2\eta_N v^2\xi}\right]$ and $\eta_N=\eta \mu(\eta)$ (or equivalently $\eta_N$ is solution of $\eta_N\nu(\eta_N)=\eta$). As mentioned by \citet{milgrom:2009vn}, the term -1 in $\nu-1$ can be replaced by $-\nu\left[\sqrt{\nu^2_N+v^4}\right]$ or by $-\nu\left[|\eta_N^2\pm v^2|\right]$ to improve the numerical convergence of the integral.

In the case of the Bekenstein approach~(\ref{eq:MONDBek}), the above integral leads only to an approximate  value for the $q$ parameter. In this approach, the $q$ parameter can only be computed by numerically solving  the non-linear Poisson equation as done in~\citep{milgrom:2009vn,blanchet:2011ys}.

From an observational point of view, the modification of the Newtonian potential (\ref{eq:EFE_SS}) will modify the trajectories of planets, asteroids and comets~\citep{milgrom:2009vn,blanchet:2011ys,blanchet:2011zr,hees:2012fk,maquet:2015jk}. Using 9 years of Cassini range and Doppler tracking measurements, the value of the parameter $Q_2$, for an external field assumed to point towards the Galactic enter, has been estimated by~\citep{hees:2014jk}
\begin{equation}\label{eq:Q2}
 Q_2=\left(3\pm3\right)\times 10^{-27} \textrm{ s}^{-2} \, .
\end{equation}

In the following, we will use the expression from Eq.~(\ref{eq:q}) to estimate the value of the $q$ parameter for different MOND transition functions and different values of the ratio $\eta$. Then, using the optimal value of $a_0$ obtained from the fit to galaxy rotation curves from Sec.~\ref{sec:rotation}, we estimate the value of the $Q_2$ parameter using the relation from Eq.~(\ref{eq:q_q2}). This value of $Q_2$ characterizes the Solar System deviation from Newtonian gravity predicted by MOND for values of $a_0$ that optimally explain galactic rotation curves. Finally, the obtained value of $Q_2$ can be compared to the Cassini estimations from \citep{hees:2014jk} to assess what transition functions are compatible with galactic rotation curves and with Solar System observations simultaneously. 

First of all, we have reproduced Tab.~I from \citep{milgrom:2009vn} to validate our calculation of $q$ using Eq.~(\ref{eq:q}). Then, we have computed $q$ for a wide range of MOND transition functions $\nu$ and values of $\eta$. The corresponding results are shown in Tab.~\ref{tab:EFE_SS_all} in the Appendix. 

Our main result consists of a combined analysis using both galactic and Solar System observations and is presented in Tab.~\ref{tab:efe_SS}. For different MOND transition functions $\nu$, the optimal MOND acceleration scale $a_0$ has been estimated with galactic rotation curves using the procedure described in Sec.~\ref{sec:rotation} (see the second column from Tab.~\ref{tab:efe_SS}). The reduced chi-square obtained for the global fit of all galactic rotation curves is also presented. Then, using the optimal value of $a_0$, we have computed the value of $Q_2$ using Eqs.(\ref{eq:q}) and~(\ref{eq:q_q2}). This estimation of $Q_2$ has been done using two different values of the external gravitational field. The two values of $g_e$ used correspond to current estimations of galactic parameters~\citep{mcmillan:2010jk,mcmillan:2011vn}: $g_e=1.9\times 10^{-10} \textrm{ m/s}^2$ and $g_e=2.4\times 10^{-10} \textrm{ m/s}^2$. The estimated values of $Q_2$ are exact in the framework of QUMOND but are only approximate estimations in the framework of the Bekenstein approach. Note that, as can be seen from Tab.~I from \citep{milgrom:2009vn}, the values obtained with the formulas from QUMOND slightly underestimate the corresponding values of $Q_2$ in the Bekenstein approach. The results from Tab.~\ref{tab:efe_SS} are therefore otpimistic in the Bekenstein approach. The estimated values of $Q_2$ presented in Tab.~\ref{tab:efe_SS} can be compared to the estimation (\ref{eq:Q2}) obtained with the Cassini radioscience tracking data~\citep{hees:2014jk}. The values of $Q_2$ within the 1$\sigma$ estimation are mentioned in boldface in Tab.~\ref{tab:efe_SS}.

Several conclusions may be drawn from this combined analysis. First of all, the class of transition functions $\tilde \nu_\alpha$ seems to be completely excluded by this combined analysis. The functions $\nu_\alpha$ and $\hat \nu_\alpha$ are excluded for low values of $\alpha$ but begins to be marginally acceptable for large values of $\alpha$. The only class of functions that seem to be able to produce a satisfactory fit to the galactic rotation curves without producing a too large deviation in the Solar System is $\bar \nu_\alpha$ for $\alpha\geq2$.

\begin{table}
 \caption{Col. 2: optimal value of the MOND acceleration scale $a_0$ found by rotation curve fits. Col. 3: reduced $\chi^2$ computed on all 27 galactic rotation curves presented in Sec.~\ref{sec:rotation}. Col. 4: value of the ratio $\eta=g_e/a_0$ with $g_e=1.9\times 10^{-10} \textrm{ m/s}^2$. Col. 5: value of $-q$ computed with Eq.~(\ref{eq:q}) for the value of $\eta$ from Col. 4. Col. 6: value of $Q_2$ obtained using Eq.~(\ref{eq:q_q2}) and the value of $-q$ from Col. 5. Col. 7: value of $\eta$ for $g_e=2.4\times 10^{-10} \textrm{ m/s}^2$. Col. 8: value of $-q$ computed with Eq.~(\ref{eq:q}) for the value of $\eta$ from Col. 7. Col. 9: value of $Q_2$ obtained using Eq.~(\ref{eq:q_q2}) and the value of $-q$ from Col. 8. The values of $Q_2$ that are in bold are included in the 1$\sigma$ Cassini estimation $0\leq Q_2\leq 6\times 10^{-27} \textrm{s}^{-2}$.}
 \label{tab:efe_SS}
 $
 \begin{array}{r|rr|rrr|rrr|}
  &      &   & \multicolumn{3}{c|}{g_e=g_{e\textrm{min}}} & \multicolumn{3}{c|}{g_e=g_{e\textrm{max}}}\\
   & a_0 & \chi^2_\textrm{red} &  \eta & -q & Q_2 &  \eta & -q & Q_2\\
  &10^{-10}&&&10^{-2}&10^{-27}&&10^{-2}&10^{-27}\\
  &[\textrm{m/s}^2]&&&&[\textrm{s}^{-2}]&&&[\textrm{s}^{-2}] \\\hline
 \nu _2 & 1.60 & 2.02 & 1.20 & 10.  & 26. & 1.50 & 11.3 & 30. \\
 \nu _3 & 1.55 & 1.97 & 1.20 & 8.29 & 21. & 1.50 & 7.82 & 20. \\
 \nu _4 & 1.51 & 1.94 & 1.30 & 6.76 & 16. & 1.60 & 5.34 & 13. \\
 \nu _5 & 1.49 & 1.93 & 1.30 & 5.51 & 13. & 1.60 & 3.71 & 8.7 \\
 \nu _6 & 1.46 & 1.92 & 1.30 & 4.55 & 11. & 1.60 & 2.67 & 6.2 \\
 \mathbf{\nu_7} & 1.45 & 1.92 & 1.30 & 3.82 & 8.7 & 1.70 & 2.01 & \textbf{4.6} \\
 \mathbf{\nu_8} & 1.44 & 1.92 & 1.30 & 3.27 & 7.3 & 1.70 & 1.58 & \textbf{3.5} \\\hline
  \tilde{\nu }_{0.5} & 1.48 & 2.16 & 1.30 & 14.8 & 35. & 1.60 & 18.5 & 44. \\
 \tilde{\nu }_{1}    & 1.38 & 2.12 &1.40 & 18.3 & 38. & 1.70 & 25. & 53. \\
 \tilde{\nu }_{1.5} & 1.18  & 2.16 & 1.60 & 24.1 & 40. & 2.00 & 34.2 & 57. \\
 \tilde{\nu }_{2} & 0.815   & 2.24 & 2.30 & 44.8 & 43. & 2.90 & 47.9 & 46. \\
 \tilde{\nu }_{2.5} & 0.977 & 2.23 & 1.90 & 33.1 & 42. & 2.50 & 51.7 & 65. \\
 \tilde{\nu }_{3} & 0.743   & 1.07 & 2.60 & 56.8 & 47. & 3.20 & 65.5 & 55. \\
 \tilde{\nu }_{4} & 0.723   & 2.01 & 2.60 & 54.8 & 44. & 3.30 & 85.9 & 69. \\
 \tilde{\nu }_{5} & 0.715   & 1.97 & 2.70 & 48.1 & 38. & 3.40 & 94.7 & 75. \\\hline
 \bar{\nu }_{0.5} & 1.48 & 2.15 & 1.30 & 13.1 & 31. & 1.60 & 17.5 & 41. \\
  \bar{\nu }_{1} & 1.38 & 2.12 &1.40 & 16.1 & 34. & 1.70 & 19.5 & 41. \\
 \bar{\nu }_{1.5} & 1.18& 2.16 & 1.60 & 19.3 & 32. & 2.00 & 15.8 & 26.5 \\
 \mathbf{\bar{\nu }_{2}} & 0.815 & 2.24 & 2.30 & 6.2 & \textbf{5.9} & 2.90 & 2.63 & \textbf{2.52} \\
 \mathbf{\bar{\nu }_{3}} & 0.743 & 2.07 & 2.60 & 1.9 & \textbf{1.6} & 3.20 & 0.82 & \textbf{0.68} \\
 \mathbf{\bar{\nu }_{4}} & 0.723 & 2.01 & 2.60 & 1.3 & \textbf{1.} & 3.30 & 0.56 & \textbf{0.45} \\
 \mathbf{\bar{\nu }_{5}} & 0.715 & 1.97 & 2.70 & 1.08 & \textbf{0.85} & 3.40 & 0. & \textbf{0.} \\
 \mathbf{\bar{\nu }_{6}} & 0.713 & 1.95 & 2.70 & 1.02 & \textbf{0.8} & 3.40 & 0. & \textbf{0.} \\
 \mathbf{\bar{\nu }_{7}} & 0.729 & 1.95 & 2.60 & 1.07 & \textbf{0.87} & 3.30 & 0. & \textbf{0.} \\\hline
 \hat{\nu }_{1} & 1.48 & 2.15 & 1.30 & 13.1 & 31. & 1.60 & 17.5 & 41. \\
 \hat{\nu }_{2} & 1.59 & 2.01 & 1.20 & 10.2 & 27. & 1.50 & 11.4 & 30. \\
 \hat{\nu }_{3} & 1.55 & 1.96 & 1.20 & 8.32 & 21. & 1.60 & 7.49 & 19. \\
 \hat{\nu }_{4} & 1.51 & 1.94 & 1.30 & 6.66 & 16. & 1.60 & 4.79 & 12. \\
 \hat{\nu }_{5} & 1.48 & 1.93 & 1.30 & 5.34 & 13. & 1.60 & 3.1 & 7.3 \\
 \mathbf{\hat{\nu }_{6}} & 1.46 & 1.92 & 1.30 & 4.31 & 9.9 & 1.60 & 2.11 & \textbf{4.9} \\
 \mathbf{\hat{\nu }_{7}} & 1.45 & 1.92 & 1.30 & 3.55 & 8.  & 1.70 & 1.55 & \textbf{3.5} \\\hline
 \end{array}
 $
\end{table}

\subsection{Prospect for the LISA pathfinder mission}
\citet{bekenstein:2006sf,bevis:2010nr,trenkel:2012qf,magueijo:2012xy,trenkel:2014nx} have proposed to redirect the Laser Interferometer Space Antenna (LISA) pathfinder towards the Earth-Sun saddle point to constrain MOND in a low gravitational field. The LISA pathfinder project~\citep{mcnamara:2008nr} is a space mission designed to test the technology to be used in the eLISA project. This mission allows the very accurate  measurement of tidal stresses by measuring the relative motion of two test masses separated by 35 cm. The idea proposed by, e.g., \citet{bevis:2010nr,trenkel:2012qf,magueijo:2012xy} is to measure the tidal stresses very close to the saddle point where the Newtonian gravitational field is very low and where MOND effects are expected to show up. In this section, we will assess the order of magnitude of the tidal stresses produced by the MOND transition functions used in the previous section and show that they are far too small to be detected by LISA pathfinder even in the most optimistic scenario.

The simulations that have been performed use similar assumptions as in \citet{magueijo:2012xy}. The spacecraft is supposed to move along the $X$ axis, which is defined by the Sun-Earth direction. Moreover, it is assumed that the direction of the observed tidal stress would be perpendicular to this axis. In order words, the anomalous observed tidal stress produced by MOND is given by $S_{yy}$ with 
\begin{equation}
 S_{ij}=\frac{\partial^2 \Phi}{\partial x^i \partial x^j}-\frac{\partial^2 \Phi_N}{\partial x^i \partial x^j}\, ,
\end{equation}
where $\Phi$ represents the MOND gravitational potential and $\Phi_N$ the Newtonian potential. Moreover, we will assume a case where the spacecraft misses the saddle point by 1 kilometer. This situation is very optimistic since as mentioned by~\citet{magueijo:2012xy}, the saddle point can be pinpointed to about a kilometer and the spacecraft location can be determined to about 10 kilometers.

\begin{figure}
\includegraphics[width=.9\columnwidth]{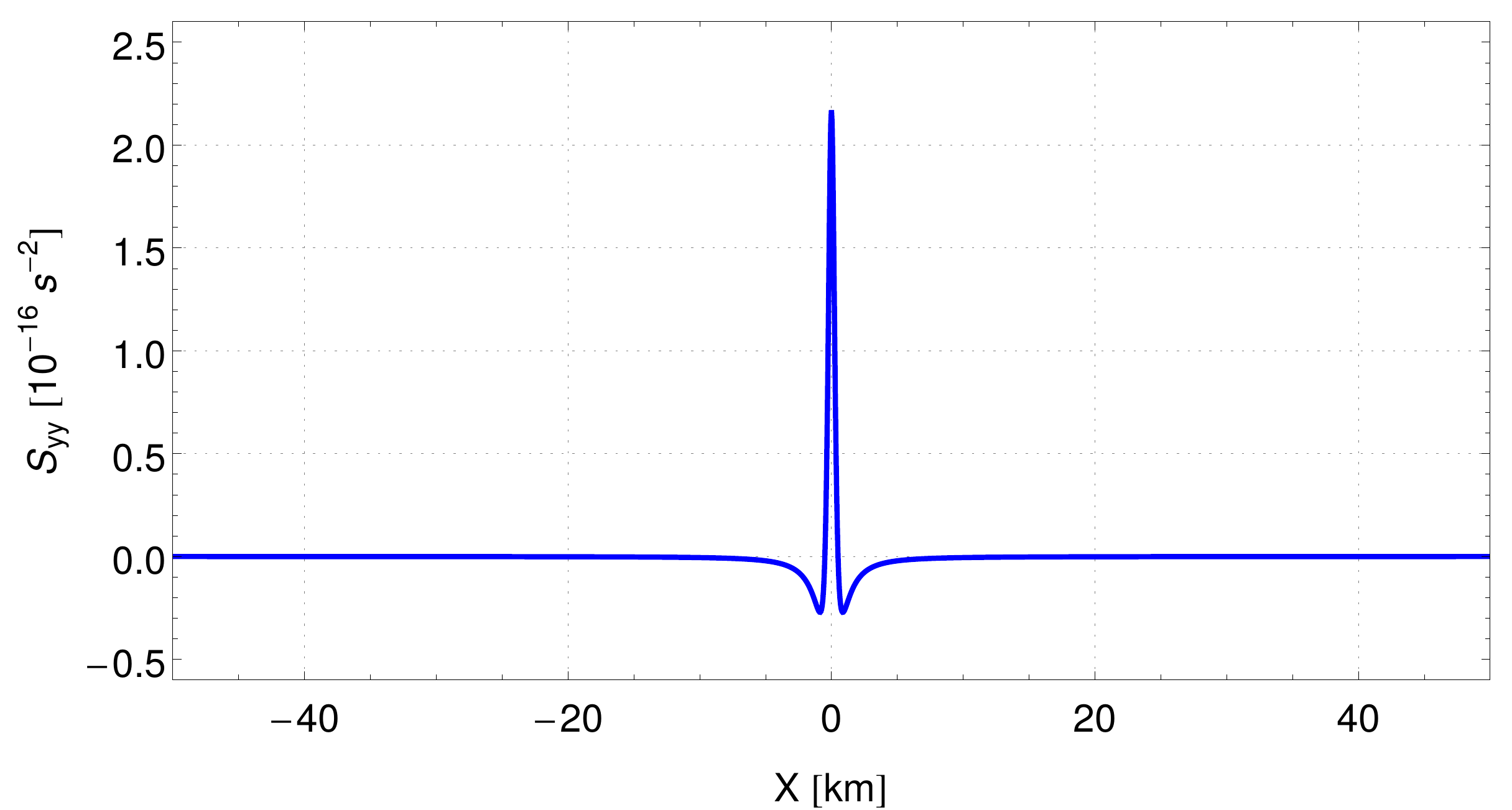}
\caption{Signature of the transverse MOND tidal stress produced by the MOND transition function $\nu_2$ (see Eq.~(\ref{eq:nun})). In this simulation, the spacecraft misses the saddle point by 1 kilometer (the origin of the $X$ axis corresponding to the closest approach with the saddle point).}
\label{fig:LISAP}
\end{figure}

Fig.~\ref{fig:LISAP} represents the evolution of the MOND tidal stress $S_{yy}$ produced by the MOND transition function $\nu_2$ (see Eq.~(\ref{eq:nun})). In this simulation, the impact factor with respect to the saddle point is 1 kilometer. The maximal amplitude of the absolute value of $S_{yy}$ is of the order of $10^{-16} \textrm{s}^{-2}$. The accuracy of LISA Pathfinder is expected to be of the order of $10^{-14} \textrm{s}^{-2}$~\citep{magueijo:2012xy}. Therefore, LISA Pathfinder will not be able to detect this MOND transition function. One might argue that the internal self-gravity of the spacecraft might however increase the effect~\citep{trenkel:2014nx}. However, let us remember that $\nu_2$ is actually excluded by our present analysis. The situation is actually much worse for the other functions: the signal for $\nu_3$ is two orders of magnitude smaller than then one produce by $\nu_2$ while the one for $\nu_4$ is 4 orders of magnitude smaller. Also recall that, for that $\nu_\alpha$ family, only $\alpha>6$ is acceptable. For other families, the function $\tilde \nu_{0.5}$ for instance produces a deviation of the order of $10^{-1000} \textrm{s}^{-2}$, while the other transition functions $\tilde \nu_\alpha$, $\bar \nu_\alpha$ and $\hat\nu_\alpha$ lead to even smaller tidal stresses. These very small numbers reflect the exponential convergence towards the Newtonian regime provided by these transition functions. Even taking into account the internal self-gravity of the spacecraft as in \citet{trenkel:2014nx} will thus not provide the necessary correction of literally multiple thousands of orders of magnitude for making the effect detectable with an acceptable transition function such as $\bar \nu_2$.

In conclusion, LISA pathfinder does not offer any possibility to constrain the transition functions considered in this analysis. The Cassini constraint from~\citet{hees:2014jk} using the External Field Effect is much more efficient. 

\section{Discussion and Conclusion}\label{sec:conclusion}
The non-linearity inherent to the MOND paradigm leads to the fact that the internal dynamics of a system is influenced by the external gravitational field in which it is embedded. In this communication, we use this EFE to derive constraints on the various MOND transition functions with a combined analysis of galactic rotation curves and of the Solar System.

First of all, we have derived the best-fit value of $a_0$ for a large class of transition functions, and we have shown that, at the galactic level, the EFE can lead to a velocity decrease in the external part of the rotation curves. This helps to improve several galactic rotation curves in our analyzed dataset, the most impressive being UGC 4173, UGC 7577 and UGC 12060.
The typical range of optimal values for the external gravitational field (ranging between $10^{-11}$ and $10^{-13}$ m/s$^2$, see Tab.~\ref{tab:resbarnu2}) is a priori realistic. It will be extremely interesting to investigate whether a source of non-negligible external field can be found in the environment of these galaxies. Nevertheless, it is not a trivial task because a massive source at large distance can contribute more than a low mass one at close distance. This also depends on the MOND cosmology \citep[e.g.][]{blanchet:2008fv,blanchet:2009kx,angus:2013jk}. For instance an external field of $10^{-12}$ m/s$^2$ can be produced by a $7\times 10^{10}M_{\sun}$ galaxy at a distance of 100 kpc, by a $3\times 10^{13}M_{\sun}$ group/cluster at 2 Mpc or by a large attractor of $2\times 10^{16}M_{\sun}$ at  50 Mpc (the typical distance from the Great Attractor to the Milky way). For instance, in the case of UGC 7577, we note that there are $\sim$ 50 galaxies at a projected distance of less than 40 kpc, which is roughly enough to produce an external field effect of $10^{-12}$ m/s$^2$ (see the estimated value from Tab.~\ref{tab:resbarnu2}, also shown in Fig.~\ref{fig:conf}).

In the Solar System, the EFE produces non negligible effects even for transition functions that present an exponential transition towards the Newtonian regime. This allows us to test MOND in the Solar System as mentioned by \citet{milgrom:2009vn,blanchet:2011ys}. Cassini observations have provided the estimation (\ref{eq:Q2}). We have performed a combined analysis of a sample of  galactic rotation curves and of the Cassini estimation to constrain the MOND transition function. The galactic rotation curves provide an estimation of the MOND acceleration scale $a_0$ that is used to estimate the $Q_2$ parameters. This estimation is compared with the observational estimation of $Q_2$ provided by Cassini data~\citep{hees:2014jk}. The results are presented in Tab.~\ref{tab:efe_SS}. The functions $\tilde \nu_\alpha$ are completely rejected by this analysis. The transition functions $\nu_\alpha$ and $\hat\nu_\alpha$ can, on the other hand, still be viable for large values of $\alpha$. The only class of functions that is compatible with both types of observations for almost all $\alpha$ is $\bar\nu_\alpha$, for $\alpha\geq 2$. We note however that these constraints do not apply to, e.g., modified inertia theories.

Finally, we have shown that for these classes of acceptable transition functions, the space mission LISA pathfinder will not be able to detect or to constrain them.

\section*{acknowledgements}
A.H. acknowledges support from ``Fonds Sp\'ecial de Recherche" through a FSR-UCL grant. We are grateful to the authors of SSM10 for sharing their data, and we acknowledge insightful discussions about the present work with Stacy McGaugh. We also acknowledge interesting discussions with Michele Armano about LISA pathfinder.

\bibliographystyle{mnras}
\bibliography{EFE}

\begin{thebibliography}{}
\makeatletter
\relax
\def\mn@urlcharsother{\let\do\@makeother \do\$\do\&\do\#\do\^\do\_\do\%\do\~}
\def\mn@doi{\begingroup\mn@urlcharsother \@ifnextchar [ {\mn@doi@}
  {\mn@doi@[]}}
\def\mn@doi@[#1]#2{\def\@tempa{#1}\ifx\@tempa\@empty \href
  {http://dx.doi.org/#2} {doi:#2}\else \href {http://dx.doi.org/#2} {#1}\fi
  \endgroup}
\def\mn@eprint#1#2{\mn@eprint@#1:#2::\@nil}
\def\mn@eprint@arXiv#1{\href {http://arxiv.org/abs/#1} {{\tt arXiv:#1}}}
\def\mn@eprint@dblp#1{\href {http://dblp.uni-trier.de/rec/bibtex/#1.xml}
  {dblp:#1}}
\def\mn@eprint@#1:#2:#3:#4\@nil{\def\@tempa {#1}\def\@tempb {#2}\def\@tempc
  {#3}\ifx \@tempc \@empty \let \@tempc \@tempb \let \@tempb \@tempa \fi \ifx
  \@tempb \@empty \def\@tempb {arXiv}\fi \@ifundefined
  {mn@eprint@\@tempb}{\@tempb:\@tempc}{\expandafter \expandafter \csname
  mn@eprint@\@tempb\endcsname \expandafter{\@tempc}}}

\bibitem[\protect\citeauthoryear{{Angus}, {van der Heyden}, {Famaey},
  {Gentile}, {McGaugh}  \& {de Blok}}{{Angus} et~al.}{2012}]{angus:2012cs}
{Angus} G.~W.,  {van der Heyden} K.~J.,  {Famaey} B.,  {Gentile} G.,  {McGaugh}
  S.~S.,   {de Blok} W.~J.~G.,  2012, \mn@doi [\mnras]
  {10.1111/j.1365-2966.2012.20532.x}, \href
  {http://adsabs.harvard.edu/abs/2012MNRAS.421.2598A} {421, 2598}

\bibitem[\protect\citeauthoryear{{Angus}, {Diaferio}, {Famaey}  \& {van der
  Heyden}}{{Angus} et~al.}{2013}]{angus:2013jk}
{Angus} G.~W.,  {Diaferio} A.,  {Famaey} B.,   {van der Heyden} K.~J.,  2013,
  \mn@doi [\mnras] {10.1093/mnras/stt1564}, \href
  {http://adsabs.harvard.edu/abs/2013MNRAS.436..202A} {436, 202}

\bibitem[\protect\citeauthoryear{{Begeman}, {Broeils}  \& {Sanders}}{{Begeman}
  et~al.}{1991}]{begeman:1991fk}
{Begeman} K.~G.,  {Broeils} A.~H.,   {Sanders} R.~H.,  1991, \mnras, \href
  {http://adsabs.harvard.edu/abs/1991MNRAS.249..523B} {249, 523}

\bibitem[\protect\citeauthoryear{{Bekenstein}}{{Bekenstein}}{2004}]{bekenstein:2004fk}
{Bekenstein} J.~D.,  2004, \mn@doi [Phys. Rev. D] {10.1103/PhysRevD.70.083509},
  70, 083509

\bibitem[\protect\citeauthoryear{{Bekenstein} \& {Magueijo}}{{Bekenstein} \&
  {Magueijo}}{2006}]{bekenstein:2006sf}
{Bekenstein} J.,  {Magueijo} J.,  2006, \mn@doi [\prd]
  {10.1103/PhysRevD.73.103513}, \href
  {http://adsabs.harvard.edu/abs/2006PhRvD..73j3513B} {73, 103513}

\bibitem[\protect\citeauthoryear{{Bekenstein} \& {Milgrom}}{{Bekenstein} \&
  {Milgrom}}{1984}]{bekenstein:1984kx}
{Bekenstein} J.,  {Milgrom} M.,  1984, \mn@doi [\apj] {10.1086/162570}, \href
  {http://adsabs.harvard.edu/abs/1984ApJ...286....7B} {286, 7}

\bibitem[\protect\citeauthoryear{{Berezhiani} \& {Khoury}}{{Berezhiani} \&
  {Khoury}}{2015}]{berezhiani:2015bh}
{Berezhiani} L.,  {Khoury} J.,  2015, preprint, \href
  {http://adsabs.harvard.edu/abs/2015arXiv150701019B} {} (\mn@eprint {arXiv}
  {1507.01019})

\bibitem[\protect\citeauthoryear{{Bernard} \& {Blanchet}}{{Bernard} \&
  {Blanchet}}{2015}]{bernard:2015sf}
{Bernard} L.,  {Blanchet} L.,  2015, \mn@doi [\prd]
  {10.1103/PhysRevD.91.103536}, \href
  {http://adsabs.harvard.edu/abs/2015PhRvD..91j3536B} {91, 103536}

\bibitem[\protect\citeauthoryear{{Bevis}, {Magueijo}, {Trenkel}  \&
  {Kemble}}{{Bevis} et~al.}{2010}]{bevis:2010nr}
{Bevis} N.,  {Magueijo} J.,  {Trenkel} C.,   {Kemble} S.,  2010, \mn@doi
  [Classical and Quantum Gravity] {10.1088/0264-9381/27/21/215014}, \href
  {http://adsabs.harvard.edu/abs/2010CQGra..27u5014B} {27, 215014}

\bibitem[\protect\citeauthoryear{{Blanchet}}{{Blanchet}}{2007}]{blanchet:2007uq}
{Blanchet} L.,  2007, \mn@doi [Classical and Quantum Gravity]
  {10.1088/0264-9381/24/14/001}, 24, 3529

\bibitem[\protect\citeauthoryear{{Blanchet} \& {Heisenberg}}{{Blanchet} \&
  {Heisenberg}}{2015}]{blanchet:2015rm}
{Blanchet} L.,  {Heisenberg} L.,  2015, \mn@doi [\prd]
  {10.1103/PhysRevD.91.103518}, \href
  {http://adsabs.harvard.edu/abs/2015PhRvD..91j3518B} {91, 103518}

\bibitem[\protect\citeauthoryear{{Blanchet} \& {Le Tiec}}{{Blanchet} \& {Le
  Tiec}}{2008}]{blanchet:2008fv}
{Blanchet} L.,  {Le Tiec} A.,  2008, \mn@doi [Phys. Rev. D]
  {10.1103/PhysRevD.78.024031}, 78, 024031

\bibitem[\protect\citeauthoryear{{Blanchet} \& {Le Tiec}}{{Blanchet} \& {Le
  Tiec}}{2009}]{blanchet:2009kx}
{Blanchet} L.,  {Le Tiec} A.,  2009, \mn@doi [Phys. Rev. D]
  {10.1103/PhysRevD.80.023524}, 80, 023524

\bibitem[\protect\citeauthoryear{{Blanchet} \& {Novak}}{{Blanchet} \&
  {Novak}}{2011a}]{blanchet:2011zr}
{Blanchet} L.,  {Novak} J.,  2011a, in {Aug\'e} E.,  {Dumarchez} J.,   {Tr\^an
  Thanh V\^an} J.,  eds, Proceedings of the XLVIth Rencontres de Moriond and
  GPhys Colloquium 2011: Gravitational Waves and Experimental Gravity. Th\^e
  Gi\'oi Publishers, Vietnam, p.~295 (\mn@eprint {arXiv} {1105.5815})

\bibitem[\protect\citeauthoryear{{Blanchet} \& {Novak}}{{Blanchet} \&
  {Novak}}{2011b}]{blanchet:2011ys}
{Blanchet} L.,  {Novak} J.,  2011b, \mn@doi [MNRAS]
  {10.1111/j.1365-2966.2010.18076.x}, 412, 2530

\bibitem[\protect\citeauthoryear{{Boylan-Kolchin}, {Bullock}  \&
  {Kaplinghat}}{{Boylan-Kolchin} et~al.}{2011}]{boylan-kolchin:2011kq}
{Boylan-Kolchin} M.,  {Bullock} J.~S.,   {Kaplinghat} M.,  2011, \mn@doi
  [\mnras] {10.1111/j.1745-3933.2011.01074.x}, \href
  {http://adsabs.harvard.edu/abs/2011MNRAS.415L..40B} {415, L40}

\bibitem[\protect\citeauthoryear{{Bruneton} \&
  {Esposito-Far{\`e}se}}{{Bruneton} \&
  {Esposito-Far{\`e}se}}{2007}]{bruneton:2007vn}
{Bruneton} J.-P.,  {Esposito-Far{\`e}se} G.,  2007, \mn@doi [Phys. Rev. D]
  {10.1103/PhysRevD.76.124012}, 76, 124012

\bibitem[\protect\citeauthoryear{{Deffayet}, {Esposito-Far{\`e}se}  \&
  {Woodard}}{{Deffayet} et~al.}{2014}]{deffayet:2014kx}
{Deffayet} C.,  {Esposito-Far{\`e}se} G.,   {Woodard} R.~P.,  2014, \mn@doi
  [\prd] {10.1103/PhysRevD.90.064038}, \href
  {http://adsabs.harvard.edu/abs/2014PhRvD..90f4038D} {90, 064038}

\bibitem[\protect\citeauthoryear{{Famaey} \& {Binney}}{{Famaey} \&
  {Binney}}{2005}]{famaey:2005rq}
{Famaey} B.,  {Binney} J.,  2005, \mn@doi [MNRAS]
  {10.1111/j.1365-2966.2005.09474.x}, 363, 603

\bibitem[\protect\citeauthoryear{{Famaey} \& {McGaugh}}{{Famaey} \&
  {McGaugh}}{2012}]{famaey:2012fk}
{Famaey} B.,  {McGaugh} S.~S.,  2012, Living Reviews in Relativity, 15, 10

\bibitem[\protect\citeauthoryear{{Famaey}, {Bruneton}  \& {Zhao}}{{Famaey}
  et~al.}{2007}]{famaey:2007dq}
{Famaey} B.,  {Bruneton} J.-P.,   {Zhao} H.,  2007, \mn@doi [\mnras]
  {10.1111/j.1745-3933.2007.00308.x}, \href
  {http://adsabs.harvard.edu/abs/2007MNRAS.377L..79F} {377, L79}

\bibitem[\protect\citeauthoryear{{Felten}}{{Felten}}{1984}]{felten:1984uq}
{Felten} J.~E.,  1984, \mn@doi [\apj] {10.1086/162569}, \href
  {http://adsabs.harvard.edu/abs/1984ApJ...286....3F} {286, 3}

\bibitem[\protect\citeauthoryear{{Gentile}, {Famaey}  \& {de Blok}}{{Gentile}
  et~al.}{2011}]{gentile:2011uq}
{Gentile} G.,  {Famaey} B.,   {de Blok} W.~J.~G.,  2011, \mn@doi [\aap]
  {10.1051/0004-6361/201015283}, \href
  {http://adsabs.harvard.edu/abs/2011A%26A...527A..76G} {527, A76}

\bibitem[\protect\citeauthoryear{{Gregory}}{{Gregory}}{2010}]{gregory:2010qv}
{Gregory} P.,  2010, {Bayesian Logical Data Analysis for the Physical Sciences}

\bibitem[\protect\citeauthoryear{{Hees} et~al.,}{{Hees}
  et~al.}{2012}]{hees:2012fk}
{Hees} A.,  et~al., 2012, \mn@doi [{Classical and Quantum Gravity}]
  {10.1088/0264-9381/29/23/235027}, 29, 235027

\bibitem[\protect\citeauthoryear{{Hees}, {Folkner}, {Jacobson}  \&
  {Park}}{{Hees} et~al.}{2014}]{hees:2014jk}
{Hees} A.,  {Folkner} W.~M.,  {Jacobson} R.~A.,   {Park} R.~S.,  2014, \mn@doi
  [\prd] {10.1103/PhysRevD.89.102002}, \href
  {http://adsabs.harvard.edu/abs/2014PhRvD..89j2002H} {89, 102002}

\bibitem[\protect\citeauthoryear{{Ibata}, {Ibata}, {Famaey}  \&
  {Lewis}}{{Ibata} et~al.}{2014}]{ibata:2014xy}
{Ibata} N.~G.,  {Ibata} R.~A.,  {Famaey} B.,   {Lewis} G.~F.,  2014, \mn@doi
  [\nat] {10.1038/nature13481}, \href
  {http://adsabs.harvard.edu/abs/2014Natur.511..563I} {511, 563}

\bibitem[\protect\citeauthoryear{{Jacobson} \& {Mattingly}}{{Jacobson} \&
  {Mattingly}}{2001}]{jacobson:2001qf}
{Jacobson} T.,  {Mattingly} D.,  2001, \mn@doi [\prd]
  {10.1103/PhysRevD.64.024028}, \href
  {http://adsabs.harvard.edu/abs/2001PhRvD..64b4028J} {64, 024028}

\bibitem[\protect\citeauthoryear{{Khoury}}{{Khoury}}{2015}]{khoury:2015qf}
{Khoury} J.,  2015, \mn@doi [\prd] {10.1103/PhysRevD.91.024022}, \href
  {http://adsabs.harvard.edu/abs/2015PhRvD..91b4022K} {91, 024022}

\bibitem[\protect\citeauthoryear{{L{\"u}ghausen}, {Famaey}  \&
  {Kroupa}}{{L{\"u}ghausen} et~al.}{2015}]{lughausen:2015jt}
{L{\"u}ghausen} F.,  {Famaey} B.,   {Kroupa} P.,  2015, \mn@doi [Canadian
  Journal of Physics] {10.1139/cjp-2014-0168}, \href
  {http://adsabs.harvard.edu/abs/2015CaJPh..93..232L} {93, 232}

\bibitem[\protect\citeauthoryear{{Magueijo} \& {Mozaffari}}{{Magueijo} \&
  {Mozaffari}}{2012}]{magueijo:2012xy}
{Magueijo} J.,  {Mozaffari} A.,  2012, \mn@doi [\prd]
  {10.1103/PhysRevD.85.043527}, \href
  {http://adsabs.harvard.edu/abs/2012PhRvD..85d3527M} {85, 043527}

\bibitem[\protect\citeauthoryear{{Maquet} \& {Pierret}}{{Maquet} \&
  {Pierret}}{2015}]{maquet:2015jk}
{Maquet} L.,  {Pierret} F.,  2015, \mn@doi [\prd] {10.1103/PhysRevD.91.084015},
  \href {http://adsabs.harvard.edu/abs/2015PhRvD..91h4015M} {91, 084015}

\bibitem[\protect\citeauthoryear{{McGaugh}}{{McGaugh}}{2012}]{mcgaugh:2012nr}
{McGaugh} S.~S.,  2012, \mn@doi [\aj] {10.1088/0004-6256/143/2/40}, \href
  {http://adsabs.harvard.edu/abs/2012AJ....143...40M} {143, 40}

\bibitem[\protect\citeauthoryear{{McMillan}}{{McMillan}}{2011}]{mcmillan:2011vn}
{McMillan} P.~J.,  2011, \mn@doi [\mnras] {10.1111/j.1365-2966.2011.18564.x},
  \href {http://adsabs.harvard.edu/abs/2011MNRAS.414.2446M} {414, 2446}

\bibitem[\protect\citeauthoryear{{McMillan} \& {Binney}}{{McMillan} \&
  {Binney}}{2010}]{mcmillan:2010jk}
{McMillan} P.~J.,  {Binney} J.~J.,  2010, \mn@doi [\mnras]
  {10.1111/j.1365-2966.2009.15932.x}, \href
  {http://adsabs.harvard.edu/abs/2010MNRAS.402..934M} {402, 934}

\bibitem[\protect\citeauthoryear{{McNamara}, {Vitale}, {Danzmann}  \& {LISA
  Pathfinder Science Working Team}}{{McNamara} et~al.}{2008}]{mcnamara:2008nr}
{McNamara} P.,  {Vitale} S.,  {Danzmann} K.,   {LISA Pathfinder Science Working
  Team} 2008, \mn@doi [Classical and Quantum Gravity]
  {10.1088/0264-9381/25/11/114034}, \href
  {http://adsabs.harvard.edu/abs/2008CQGra..25k4034M} {25, 114034}

\bibitem[\protect\citeauthoryear{{Milgrom}}{{Milgrom}}{1983a}]{milgrom:1983fk}
{Milgrom} M.,  1983a, \mn@doi [ApJ] {10.1086/161130}, 270, 365

\bibitem[\protect\citeauthoryear{{Milgrom}}{{Milgrom}}{1983b}]{milgrom:1983uq}
{Milgrom} M.,  1983b, \mn@doi [ApJ] {10.1086/161131}, 270, 371

\bibitem[\protect\citeauthoryear{{Milgrom}}{{Milgrom}}{1994}]{milgrom:1994zr}
{Milgrom} M.,  1994, \mn@doi [Annals of Physics] {10.1006/aphy.1994.1012},
  \href {http://adsabs.harvard.edu/abs/1994AnPhy.229..384M} {229, 384}

\bibitem[\protect\citeauthoryear{{Milgrom}}{{Milgrom}}{2009a}]{milgrom:2009kx}
{Milgrom} M.,  2009a, \mn@doi [\prd] {10.1103/PhysRevD.80.123536}, \href
  {http://adsabs.harvard.edu/abs/2009PhRvD..80l3536M} {80, 123536}

\bibitem[\protect\citeauthoryear{{Milgrom}}{{Milgrom}}{2009b}]{milgrom:2009vn}
{Milgrom} M.,  2009b, \mn@doi [MNRAS] {10.1111/j.1365-2966.2009.15302.x}, 399,
  474

\bibitem[\protect\citeauthoryear{{Milgrom}}{{Milgrom}}{2010}]{milgrom:2010uq}
{Milgrom} M.,  2010, \mn@doi [\mnras] {10.1111/j.1365-2966.2009.16184.x}, \href
  {http://adsabs.harvard.edu/abs/2010MNRAS.403..886M} {403, 886}

\bibitem[\protect\citeauthoryear{{Milgrom}}{{Milgrom}}{2011}]{milgrom:2011fk}
{Milgrom} M.,  2011, \mn@doi [Acta Physica Polonica B]
  {10.5506/APhysPolB.42.2175}, \href
  {http://adsabs.harvard.edu/abs/2011arXiv1111.1611M} {42, 2175}

\bibitem[\protect\citeauthoryear{{Milgrom}}{{Milgrom}}{2014}]{milgrom:2014ix}
{Milgrom} M.,  2014, \mn@doi [Scholarpedia] {10.4249/scholarpedia.31410}, \href
  {http://adsabs.harvard.edu/abs/2014SchpJ...931410M} {9, 31410}

\bibitem[\protect\citeauthoryear{{Oman} et~al.,}{{Oman}
  et~al.}{2015}]{oman:2015rm}
{Oman} K.~A.,  et~al., 2015, \mn@doi [\mnras] {10.1093/mnras/stv1504}, \href
  {http://adsabs.harvard.edu/abs/2015MNRAS.452.3650O} {452, 3650}

\bibitem[\protect\citeauthoryear{{Pawlowski}, {Pflamm-Altenburg}  \&
  {Kroupa}}{{Pawlowski} et~al.}{2012}]{pawlowski:2012qf}
{Pawlowski} M.~S.,  {Pflamm-Altenburg} J.,   {Kroupa} P.,  2012, \mn@doi
  [\mnras] {10.1111/j.1365-2966.2012.20937.x}, \href
  {http://adsabs.harvard.edu/abs/2012MNRAS.423.1109P} {423, 1109}

\bibitem[\protect\citeauthoryear{{Planck Collaboration XVI}}{{Planck
  Collaboration XVI}}{2014}]{planck-collaboration:2014yf}
{Planck Collaboration XVI} 2014, \mn@doi [\aap] {10.1051/0004-6361/201321591},
  \href {http://adsabs.harvard.edu/abs/2014A%26A...571A..16P} {571, A16}

\bibitem[\protect\citeauthoryear{{Sanders}}{{Sanders}}{1997}]{sanders:1997kx}
{Sanders} R.~H.,  1997, \mn@doi [\apj] {10.1086/303980}, \href
  {http://adsabs.harvard.edu/abs/1997ApJ...480..492S} {480, 492}

\bibitem[\protect\citeauthoryear{{Sanders}}{{Sanders}}{2005}]{sanders:2005fk}
{Sanders} R.~H.,  2005, \mn@doi [\mnras] {10.1111/j.1365-2966.2005.09375.x},
  363, 459

\bibitem[\protect\citeauthoryear{{Sanders} \& {Noordermeer}}{{Sanders} \&
  {Noordermeer}}{2007}]{sanders:2007ly}
{Sanders} R.~H.,  {Noordermeer} E.,  2007, \mn@doi [\mnras]
  {10.1111/j.1365-2966.2007.11981.x}, \href
  {http://adsabs.harvard.edu/abs/2007MNRAS.379..702S} {379, 702}

\bibitem[\protect\citeauthoryear{{Sanders} \& {Verheijen}}{{Sanders} \&
  {Verheijen}}{1998}]{sanders:1998uq}
{Sanders} R.~H.,  {Verheijen} M.~A.~W.,  1998, \mn@doi [\apj] {10.1086/305986},
  \href {http://adsabs.harvard.edu/abs/1998ApJ...503...97S} {503, 97}

\bibitem[\protect\citeauthoryear{{Swaters} \& {Balcells}}{{Swaters} \&
  {Balcells}}{2002}]{swaters:2002qf}
{Swaters} R.~A.,  {Balcells} M.,  2002, \mn@doi [\aap]
  {10.1051/0004-6361:20020449}, \href
  {http://adsabs.harvard.edu/abs/2002A%26A...390..863S} {390, 863}

\bibitem[\protect\citeauthoryear{{Swaters}, {Sanders}  \& {McGaugh}}{{Swaters}
  et~al.}{2010}]{swaters:2010eu}
{Swaters} R.~A.,  {Sanders} R.~H.,   {McGaugh} S.~S.,  2010, \mn@doi [\apj]
  {10.1088/0004-637X/718/1/380}, \href
  {http://adsabs.harvard.edu/abs/2010ApJ...718..380S} {718, 380}

\bibitem[\protect\citeauthoryear{{Trenkel} \& {Wealthy}}{{Trenkel} \&
  {Wealthy}}{2014}]{trenkel:2014nx}
{Trenkel} C.,  {Wealthy} D.,  2014, \mn@doi [\prd]
  {10.1103/PhysRevD.90.084037}, \href
  {http://adsabs.harvard.edu/abs/2014PhRvD..90h4037T} {90, 084037}

\bibitem[\protect\citeauthoryear{{Trenkel}, {Kemble}, {Bevis}  \&
  {Magueijo}}{{Trenkel} et~al.}{2012}]{trenkel:2012qf}
{Trenkel} C.,  {Kemble} S.,  {Bevis} N.,   {Magueijo} J.,  2012, \mn@doi
  [Advances in Space Research] {10.1016/j.asr.2012.07.024}, \href
  {http://adsabs.harvard.edu/abs/2012AdSpR..50.1570T} {50, 1570}

\bibitem[\protect\citeauthoryear{{Vogelsberger} et~al.,}{{Vogelsberger}
  et~al.}{2014}]{vogelsberger:2014sf}
{Vogelsberger} M.,  et~al., 2014, \mn@doi [\mnras] {10.1093/mnras/stu1536},
  \href {http://adsabs.harvard.edu/abs/2014MNRAS.444.1518V} {444, 1518}

\bibitem[\protect\citeauthoryear{{Wu}, {Famaey}, {Gentile}, {Perets}  \&
  {Zhao}}{{Wu} et~al.}{2008}]{wu:2008cr}
{Wu} X.,  {Famaey} B.,  {Gentile} G.,  {Perets} H.,   {Zhao} H.,  2008, \mn@doi
  [\mnras] {10.1111/j.1365-2966.2008.13198.x}, \href
  {http://adsabs.harvard.edu/abs/2008MNRAS.386.2199W} {386, 2199}

\bibitem[\protect\citeauthoryear{{Zhao} \& {Famaey}}{{Zhao} \&
  {Famaey}}{2006}]{zhao:2006la}
{Zhao} H.~S.,  {Famaey} B.,  2006, \mn@doi [\apjl] {10.1086/500805}, \href
  {http://adsabs.harvard.edu/abs/2006ApJ...638L...9Z} {638, L9}

\bibitem[\protect\citeauthoryear{{Zhao} \& {Famaey}}{{Zhao} \&
  {Famaey}}{2010}]{zhao:2010uq}
{Zhao} H.,  {Famaey} B.,  2010, \mn@doi [\prd] {10.1103/PhysRevD.81.087304},
  \href {http://adsabs.harvard.edu/abs/2010PhRvD..81h7304Z} {81, 087304}

\bibitem[\protect\citeauthoryear{{Zlosnik}, {Ferreira}  \&
  {Starkman}}{{Zlosnik} et~al.}{2006}]{zlosnik:2006dq}
{Zlosnik} T.~G.,  {Ferreira} P.~G.,   {Starkman} G.~D.,  2006, \mn@doi [\prd]
  {10.1103/PhysRevD.74.044037}, \href
  {http://adsabs.harvard.edu/abs/2006PhRvD..74d4037Z} {74, 044037}

\bibitem[\protect\citeauthoryear{{Zlosnik}, {Ferreira}  \&
  {Starkman}}{{Zlosnik} et~al.}{2007}]{zlosnik:2007bh}
{Zlosnik} T.~G.,  {Ferreira} P.~G.,   {Starkman} G.~D.,  2007, \mn@doi [\prd]
  {10.1103/PhysRevD.75.044017}, \href
  {http://adsabs.harvard.edu/abs/2007PhRvD..75d4017Z} {75, 044017}

\bibitem[\protect\citeauthoryear{{Zwaan}, {van der Hulst}, {de Blok}  \&
  {McGaugh}}{{Zwaan} et~al.}{1995}]{zwaan:1995yg}
{Zwaan} M.~A.,  {van der Hulst} J.~M.,  {de Blok} W.~J.~G.,   {McGaugh} S.~S.,
  1995, \mnras, \href {http://adsabs.harvard.edu/abs/1995MNRAS.273L..35Z} {273,
  L35}

\bibitem[\protect\citeauthoryear{{de Blok} \& {McGaugh}}{{de Blok} \&
  {McGaugh}}{1998}]{de-blok:1998zr}
{de Blok} W.~J.~G.,  {McGaugh} S.~S.,  1998, \mn@doi [\apj] {10.1086/306390},
  \href {http://adsabs.harvard.edu/abs/1998ApJ...508..132D} {508, 132}

\makeatother
\end{thebibliography}

\begin{appendix}
	
\section{Fit to rotation curves with $\nu_8$ and $\hat\nu_6$}\label{app:fit}

Here we present fits using the transition functions $\nu_8$ and $\hat\nu_6$ which, like $\bar\nu_2$, are not rejected by Solar System observations. It is interesting to notice that, at the level of galactic rotation curves, fits using $\nu_\alpha$ and $\hat\nu_\alpha$ are very similar. This is explained by the similarity in the profile of the transition function as can be seen in Fig.~\ref{fig:nu} and can be noticed from the first columns of Tab.~\ref{tab:efe_SS} where the optimal values for the MOND acceleration scales and the $\chi^2$ of rotation curves is presented for different transition functions. It can be seen that values for $\nu_\alpha$ and $\hat\nu_\alpha$ are very similar. 

\begin{figure*}
\includegraphics[width=1.85\columnwidth]{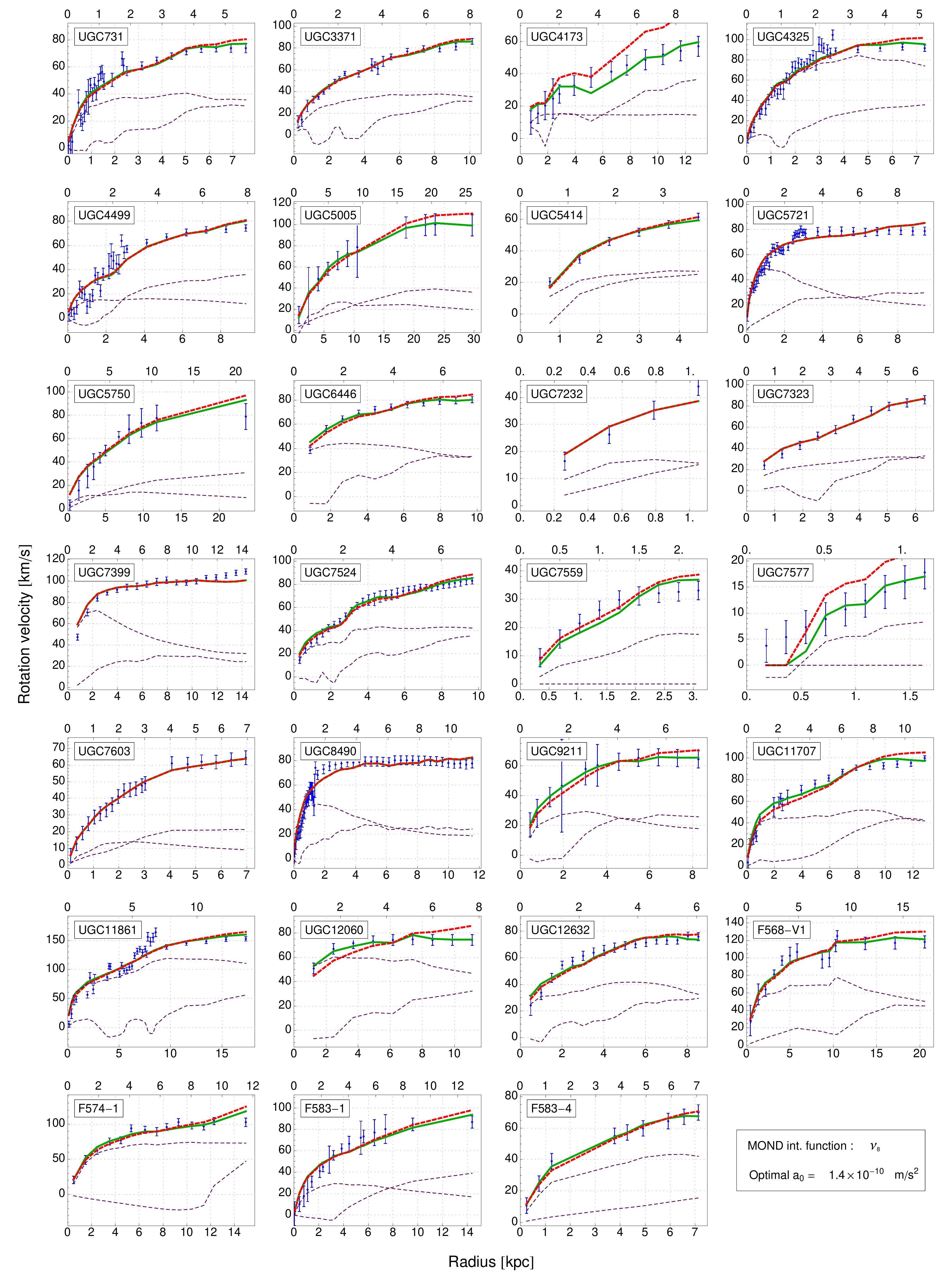}
\caption{Results of the fits using the MOND transition function $\nu_8$ for the optimal value $a_0=1.4\times 10^{-10}$ m/s$^2$. The dashed (red) thick lines represent the optimal fit without any EFE ; the thick (green) solid line represents the optimal fit with EFE ; the thin solid line represents Newtonian the contributions of the stars and the thin dashed line represents the gas contribution. Since the optimal fits with and without EFE do not necessarily produce the same distance scale factor, the radial scales may not be the same. On the top of the plots we mention the radial scale obtained without EFE, at the bottom of the plots we mention the radial scale obtained with EFE.}
\label{fig:resnu8}
\end{figure*}

\begin{figure*}
\includegraphics[width=1.85\columnwidth]{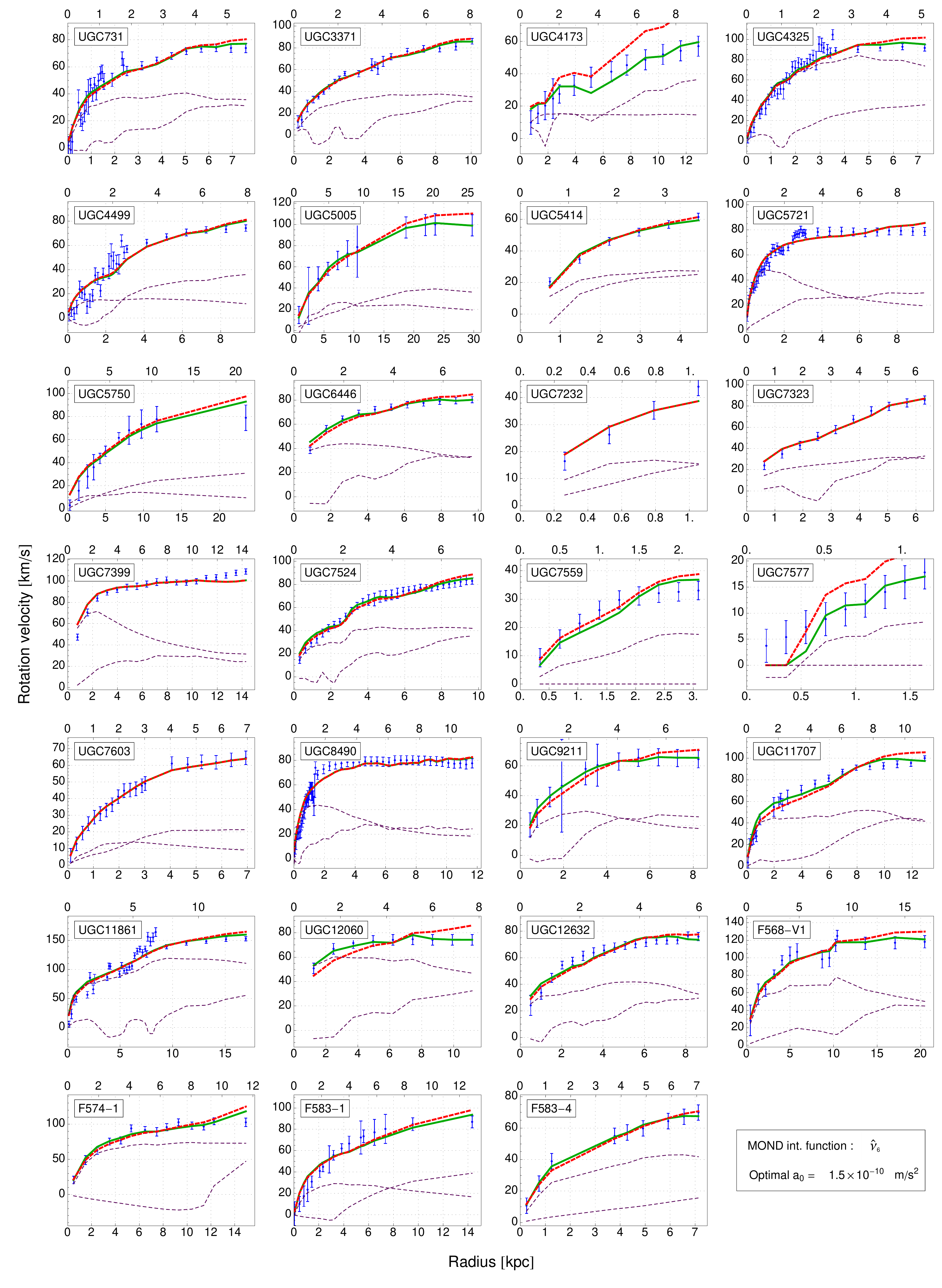}
\caption{Results of the fits using the MOND transition function $\hat\nu_6$ for the optimal value $a_0=1.5\times 10^{-10}$ m/s$^2$. The dashed (red) thick lines represent the optimal fit without any EFE ; the thick (green) solid line represents the optimal fit with EFE ; the thin solid line represents Newtonian the contributions of the stars and the thin dashed line represents the gas contribution. Since the optimal fits with and without EFE do not necessarily produce the same distance scale factor, the radial scales may not be the same. On the top of the plots we mention the radial scale obtained without EFE, atalso the bottom of the plots we mention the radial scale obtained with EFE.}
\label{fig:reshatnu6}
\end{figure*}

\begin{figure*}
\includegraphics[width=.9\columnwidth]{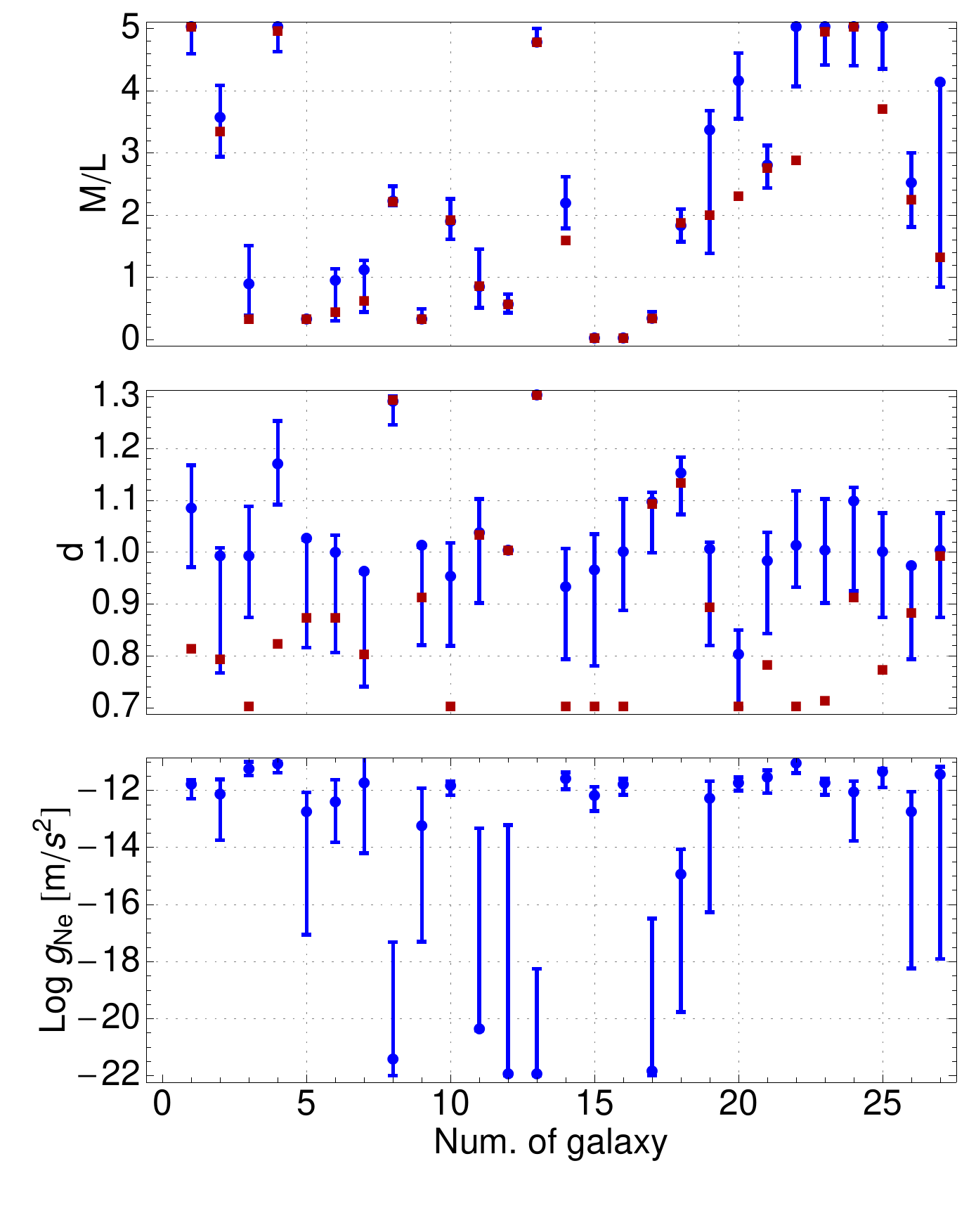}\hfill
\includegraphics[width=.9\columnwidth]{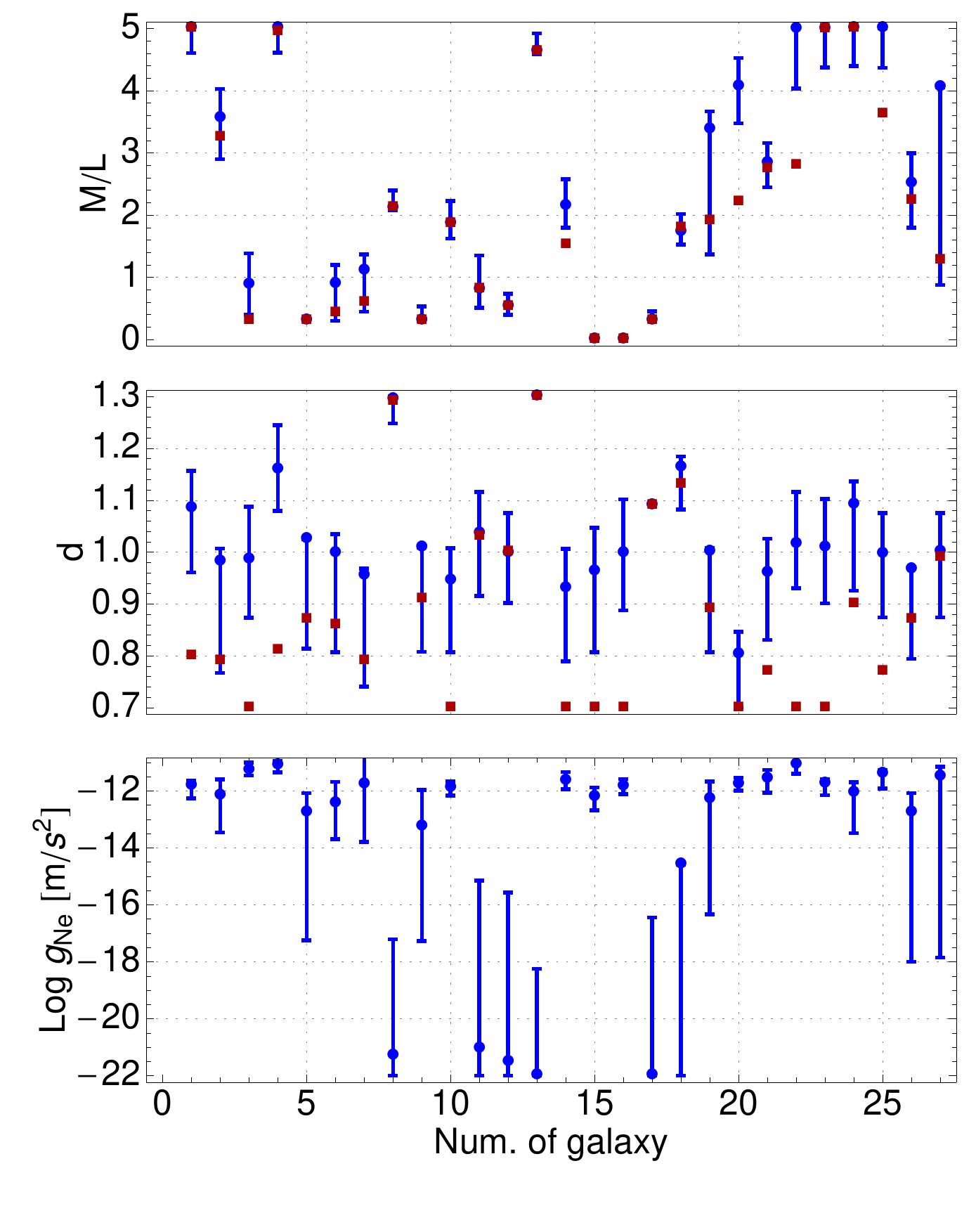}
\caption{Left: fits with $\nu_8$ and $a_0=1.4\times 10^{-10}$ m/s$^2$. Right: fits with $\hat\nu_6$ and $a_0=1.5\times 10^{-10}$ m/s$^2$. Blue: 68 \% Bayesian confidence intervals obtained with local fits to each of the 27 galaxies with the transition function. The blue dots represent the best fit. The red squares represent the optimal values obtained without any external field effect.}
\label{fig:confnuapp}
\end{figure*}

Fig.~\ref{fig:resnu8} represents the rotation curve fits for $\nu_8$ and Fig.~\ref{fig:reshatnu6} for $\hat\nu_6$. The optimal values and confidence intervals for $\Upsilon_g$, $d_g$ and the external field effects are presented in Fig.~\ref{fig:confnuapp}. The fits for both of these transition functions are qualitatively similar. The EFE improves more fits with $\nu_8$ and $\hat\nu_6$ than with $\bar \nu_2$. As with $\bar\nu_2$, the quality of the fits of UGC 4173 and UGC 11707 and UGC 7577 are significantly improved and the fit for UGC 12060 is improved for all the radii. In addition, the EFE improves the quality of the fits for UGC 731, UGC 4325, UGC 5005, UGC 6446, UGC 7524, UGC 7559, UGC 9211, F568-V1, F574-1, F583-1 and F583-4. More fits are improved using $\nu_8$ and $\hat\nu_6$ compared to $\bar\nu_2$ and for some galaxies, the improvement is  more significant as well (see for example UGC 4325).

\section{Computation of the quadrupolar EFE in the Solar System}
Here, we have computed $q$ using Eq.~(\ref{eq:q}) for a wide range of MOND transition functions $\nu$ and values of the external field $\eta$. The results are shown in Tab.~\ref{tab:EFE_SS_all}. 

\begin{table*}
\caption{Value of the parameter $q$ computed using Eq.~(\ref{eq:q}) for different MOND interpolating function $\nu$ and value of $\eta$.}
\label{tab:EFE_SS_all}
$\begin{array}{c|ccccccccc}
 \eta  & 1. & 1.25 & 1.5 & 1.75 & 2. & 2.25 & 2.5 & 2.75 & 3 \\\hline
 \nu _2 & 8.8\times 10^{-2} & 1.\times 10^{-1} & 1.1\times 10^{-1} & 1.2\times 10^{-1} & 1.2\times 10^{-1} & 1.2\times 10^{-1} &
   1.2\times 10^{-1} & 1.2\times 10^{-1} & 1.2\times 10^{-1} \\
 \nu _3 & 8.\times 10^{-2} & 8.3\times 10^{-2} & 7.9\times 10^{-2} & 7.3\times 10^{-2} & 6.7\times 10^{-2} & 6.\times 10^{-2} &
   5.4\times 10^{-2} & 4.9\times 10^{-2} & 4.5\times 10^{-2} \\
 \nu _4 & 7.4\times 10^{-2} & 6.8\times 10^{-2} & 5.7\times 10^{-2} & 4.7\times 10^{-2} & 3.8\times 10^{-2} & 3.1\times 10^{-2} &
   2.6\times 10^{-2} & 2.1\times 10^{-2} & 1.8\times 10^{-2} \\
 \nu _5 & 7.\times 10^{-2} & 5.7\times 10^{-2} & 4.3\times 10^{-2} & 3.1\times 10^{-2} & 2.3\times 10^{-2} & 1.8\times 10^{-2} &
   1.4\times 10^{-2} & 1.1\times 10^{-2} & 8.4\times 10^{-3} \\
 \nu _6 & 6.6\times 10^{-2} & 4.9\times 10^{-2} & 3.3\times 10^{-2} & 2.3\times 10^{-2} & 1.6\times 10^{-2} & 1.1\times 10^{-2} &
   8.\times 10^{-3} & 6.\times 10^{-3} & 4.5\times 10^{-3} \\
 \nu _7 & 6.3\times 10^{-2} & 4.3\times 10^{-2} & 2.7\times 10^{-2} & 1.7\times 10^{-2} & 1.1\times 10^{-2} & 7.6\times 10^{-3} &
   5.4\times 10^{-3} & 3.9\times 10^{-3} & 2.9\times 10^{-3} \\
 \nu _8 & 6.1\times 10^{-2} & 3.8\times 10^{-2} & 2.2\times 10^{-2} & 1.4\times 10^{-2} & 8.6\times 10^{-3} & 5.7\times 10^{-3} &
   4.\times 10^{-3} & 2.8\times 10^{-3} & 2.1\times 10^{-3} \\\hline
 \tilde{\nu }_{.5} & 1.1\times 10^{-1} & 1.4\times 10^{-1} & 1.7\times 10^{-1} & 2.\times 10^{-1} & 2.1\times 10^{-1} & 2.1\times
   10^{-1} & 2.\times 10^{-1} & 2.\times 10^{-1} & 1.9\times 10^{-1} \\
 \tilde{\nu }_1 & 1.1\times 10^{-1} & 1.6\times 10^{-1} & 2.1\times 10^{-1} & 2.5\times 10^{-1} & 2.8\times 10^{-1} & 3.\times
   10^{-1} & 3.\times 10^{-1} & 2.9\times 10^{-1} & 2.8\times 10^{-1} \\
 \tilde{\nu }_{1.5} & 1.\times 10^{-1} & 1.5\times 10^{-1} & 2.1\times 10^{-1} & 2.8\times 10^{-1} & 3.4\times 10^{-1} & 3.8\times
   10^{-1} & 3.9\times 10^{-1} & 3.9\times 10^{-1} & 3.7\times 10^{-1} \\
 \tilde{\nu }_2 & 9.4\times 10^{-2} & 1.4\times 10^{-1} & 2.1\times 10^{-1} & 2.8\times 10^{-1} & 3.6\times 10^{-1} & 4.3\times
   10^{-1} & 4.8\times 10^{-1} & 4.9\times 10^{-1} & 4.7\times 10^{-1} \\
 \tilde{\nu }_{2.5} & 8.7\times 10^{-2} & 1.3\times 10^{-1} & 1.9\times 10^{-1} & 2.7\times 10^{-1} & 3.5\times 10^{-1} & 4.5\times
   10^{-1} & 5.3\times 10^{-1} & 5.8\times 10^{-1} & 5.8\times 10^{-1} \\
 \tilde{\nu }_3 & 8.\times 10^{-2} & 1.2\times 10^{-1} & 1.8\times 10^{-1} & 2.5\times 10^{-1} & 3.3\times 10^{-1} & 4.3\times
   10^{-1} & 5.4\times 10^{-1} & 6.4\times 10^{-1} & 6.7\times 10^{-1} \\
 \tilde{\nu }_4 & 6.9\times 10^{-2} & 1.1\times 10^{-1} & 1.5\times 10^{-1} & 2.1\times 10^{-1} & 2.9\times 10^{-1} & 3.8\times
   10^{-1} & 4.8\times 10^{-1} & 6.2\times 10^{-1} & 7.7\times 10^{-1} \\
 \tilde{\nu }_5 & 6.1\times 10^{-2} & 9.4\times 10^{-2} & 1.4\times 10^{-1} & 1.9\times 10^{-1} & 2.5\times 10^{-1} & 3.2\times
   10^{-1} & 4.1\times 10^{-1} & 5.2\times 10^{-1} & 6.6\times 10^{-1} \\\hline%
 \bar{\nu }_1 & 1.1\times 10^{-1} & 1.4\times 10^{-1} & 1.7\times 10^{-1} & 2.\times 10^{-1} & 2.1\times 10^{-1} & 2.1\times 10^{-1}
   & 2.\times 10^{-1} & 2.\times 10^{-1} & 1.9\times 10^{-1} \\
 \bar{\nu }_{1.5} & 1.2\times 10^{-1} & 1.6\times 10^{-1} & 1.9\times 10^{-1} & 1.9\times 10^{-1} & 1.6\times 10^{-1} & 1.3\times
   10^{-1} & 1.1\times 10^{-1} & 8.3\times 10^{-2} & 6.5\times 10^{-2} \\
 \bar{\nu }_2 & 1.2\times 10^{-1} & 1.8\times 10^{-1} & 2.\times 10^{-1} & 1.5\times 10^{-1} & 1.\times 10^{-1} & 7.\times 10^{-2} &
   4.8\times 10^{-2} & 3.4\times 10^{-2} & 2.5\times 10^{-2} \\
 \bar{\nu }_3 & 1.2\times 10^{-1} & 1.8\times 10^{-1} & 1.5\times 10^{-1} & 8.1\times 10^{-2} & 4.8\times 10^{-2} & 3.1\times
   10^{-2} & 2.1\times 10^{-2} & 1.5\times 10^{-2} & 1.1\times 10^{-2} \\
 \bar{\nu }_4 & 1.1\times 10^{-1} & 1.8\times 10^{-1} & 1.1\times 10^{-1} & 5.8\times 10^{-2} & 3.5\times 10^{-2} & 2.3\times
   10^{-2} & 1.5\times 10^{-2} & 1.1\times 10^{-2} & 8.\times 10^{-3} \\
 \bar{\nu }_5 & 1.1\times 10^{-1} & 1.7\times 10^{-1} & 9.1\times 10^{-2} & 5.\times 10^{-2} & 3.1\times 10^{-2} & 2.\times 10^{-2}
   & 1.4\times 10^{-2} & 9.6\times 10^{-3} & 7.1\times 10^{-3} \\
 \bar{\nu }_6 & 1.1\times 10^{-1} & 1.7\times 10^{-1} & 8.5\times 10^{-2} & 4.7\times 10^{-2} & 2.9\times 10^{-2} & 1.9\times
   10^{-2} & 1.3\times 10^{-2} & 9.1\times 10^{-3} & 0. \\
   \bar{\nu }_7 &  1.1\times 10^{-1} & 1.7\times 10^{-1}  & 2.7\times 10^{-1} & 4.6\times
   10^{-2} & 2.8\times 10^{-2} & 1.8\times 10^{-2} & 1.2\times 10^{-2} & 0. & 0. \\\hline
 \hat{\nu }_1 & 9.5\times 10^{-2} & 1.3\times 10^{-1} & 1.6\times 10^{-1} & 1.9\times 10^{-1} & 2.2\times 10^{-1} & 2.5\times
   10^{-1} & 2.8\times 10^{-1} & 3.\times 10^{-1} & 3.3\times 10^{-1} \\
 \hat{\nu }_2 & 9.\times 10^{-2} & 1.1\times 10^{-1} & 1.1\times 10^{-1} & 1.2\times 10^{-1} & 1.2\times 10^{-1} & 1.1\times 10^{-1}
   & 1.1\times 10^{-1} & 1.\times 10^{-1} & 9.6\times 10^{-2} \\
 \hat{\nu }_3 & 8.1\times 10^{-2} & 8.3\times 10^{-2} & 7.7\times 10^{-2} & 6.7\times 10^{-2} & 5.6\times 10^{-2} & 4.6\times
   10^{-2} & 3.6\times 10^{-2} & 2.9\times 10^{-2} & 2.3\times 10^{-2} \\
 \hat{\nu }_4 & 7.5\times 10^{-2} & 6.7\times 10^{-2} & 5.3\times 10^{-2} & 3.9\times 10^{-2} & 2.8\times 10^{-2} & 1.9\times
   10^{-2} & 1.3\times 10^{-2} & 9.4\times 10^{-3} & 6.8\times 10^{-3} \\
 \hat{\nu }_5 & 7.\times 10^{-2} & 5.6\times 10^{-2} & 3.8\times 10^{-2} & 2.4\times 10^{-2} & 1.5\times 10^{-2} & 9.9\times 10^{-3}
   & 6.7\times 10^{-3} & 4.7\times 10^{-3} & 3.4\times 10^{-3} \\
 \hat{\nu }_6 & 6.6\times 10^{-2} & 4.7\times 10^{-2} & 2.9\times 10^{-2} & 1.7\times 10^{-2} & 1.\times 10^{-2} & 6.5\times 10^{-3}
   & 4.4\times 10^{-3} & 3.1\times 10^{-3} & 2.3\times 10^{-3} \\
 \hat{\nu }_7 & 6.4\times 10^{-2} & 4.1\times 10^{-2} & 2.3\times 10^{-2} & 1.3\times 10^{-2} & 7.6\times 10^{-3} & 4.9\times
   10^{-3} & 3.4\times 10^{-3} & 2.4\times 10^{-3} & 1.7\times 10^{-3} \\
\end{array}
$
\end{table*}
\end{appendix}

\label{lastpage}

\end{document}